\documentclass{emulateapj}  

\usepackage{graphicx} 
\usepackage{apjfonts}

\newcommand{\ia}{\'{\i}}

\shorttitle{Emergence of magnetic bubbles through the solar atmosphere}
\shortauthors{Ortiz et al.}

\begin{document}

\title{Emergence of granular-sized magnetic bubbles through the solar atmosphere. 
I.~Spectropolarimetric observations and simulations}

\author{Ada Ortiz} 
\affil{Institute of Theoretical Astrophysics,
  University of Oslo, P.O. Box 1029 Blindern, N-0315 Oslo, Norway}
\email{ada@astro.uio.no} 
\author{Luis Ram\'on Bellot Rubio}
\affil{Instituto de Astrof\ia sica de Andaluc\ia a (CSIC), Apdo.\
  3040, 18080 Granada, Spain} 
  \author{Viggo H. Hansteen}
\affil{Institute of Theoretical Astrophysics, University of Oslo, P.O.
  Box 1029 Blindern, N-0315 Oslo, Norway} 
\author{Jaime de la Cruz Rodr\ia guez} 
\affil{Department of Physics and Astronomy, Uppsala
  University, Box 516, SE-75120 Uppsala, Sweden} 
  \and
  \author{Luc Rouppe van der Voort} 
  \affil{Institute of Theoretical Astrophysics,
  University of Oslo, P.O. Box 1029 Blindern, N-0315 Oslo, Norway}

\begin{abstract}

  We study a granular-sized magnetic flux emergence event that
  occurred in NOAA 11024 in July 2009. The observations were made with
  the CRISP spectropolarimeter at the Swedish 1 m Solar Telescope
  achieving a spatial resolution of 0.14\arcsec. Simultaneous full
  Stokes observations of the two photospheric \ion{Fe}{1} lines at
  630.2~nm and the chromospheric \ion{Ca}{2} 854.2~nm line allow us to
  describe in detail the emergence process across the solar
  atmosphere. We report here on 3D semi-spherical bubble events, where
  instead of simple magnetic footpoints, we observe complex
  semi-circular feet straddling a few granules.  Several phenomena
  occur simultaneously, namely, abnormal granulation, separation of
  opposite-polarity legs, and brightenings at chromospheric heights.
  However, the most characteristic signature in these events is the
  observation of a {\em dark bubble} in filtergrams taken in the wings
  of the \ion{Ca}{2} 854.2~nm line. There is a clear coincidence
  between the emergence of horizontal magnetic field patches and the
  formation of the dark bubble. We can infer how the bubble rises
  through the solar atmosphere as we see it progressing from the wings
  to the core of \ion{Ca}{2} 854.2~nm. In the photosphere, the
  magnetic bubble shows mean upward Doppler velocities of
  2~km\,s$^{-1}$ and expands at a horizontal speed of 4~km\,s$^{-1}$.
  In about 3.5~minutes it travels some 1100~km to reach the mid
  chromosphere, implying an average ascent speed of 5.2~km\,s$^{-1}$.
  The maximum separation attained by the magnetic legs is 6\farcs6.
  From an inversion of the observed Stokes spectra with the SIR code
  we find maximum photospheric field strengths of 480~G and
  inclinations of nearly $90^{\circ}$ in the magnetic bubble interior,
  along with temperature deficits of up to 250~K at $\log \tau = -2$
  and above. To aid the interpretation of the observations, we carry
  out 3D numerical simulations of the evolution of a horizontal, {\bf untwisted
  magnetic flux sheet} injected in the convection zone, using the
  Bifrost code.  The computational domain spans from the upper
  convection zone to the lower corona. In the modeled chromosphere the
  rising flux {\bf sheet} produces a large, cool, magnetized bubble.
  We compare this bubble with the observed ones and find excellent
  agreement, including similar field strengths and velocity signals in
  the photosphere and chromosphere, temperature deficits, ascent
  speeds, expansion velocities, and lifetimes.

\end{abstract}


\section{Introduction}

The currently accepted picture of the active Sun portraits solar
magnetic fields being created as magnetic flux tubes in the tachocline
(the interface between the radiative interior and the differentially
rotating outer convection zone of the Sun) due to extreme shear.
Buoyancy instabilities make the flux tubes to rise as $\Omega$-shaped
loops and reach the upper convection zone, emerging into the
photosphere in the form of active regions \citep[see,
e.g.,][]{caligari:1995, moreno:1996, fisher:2000}. {\bf This scenario
  is supported by more recent simulations that include the effects of
  turbulent, solar-like convective flows on rising thin flux tubes
  \citep{weber}, 3D MHD flux tubes \citep{jouve}, and buoyant magnetic
  loops \citep{nelson}}.  The emergence occurs in a plethora of sizes
and magnetic flux contents, stretching from the largest activity
complexes, with sunspots 30000 km wide and fluxes of the order of
$10^{22}$~Mx, to pores, faculae and the smaller ephemeral regions,
with fluxes between $10^{18}$ and $10^{20}$~Mx \citep[e.g.,][]{mh79,
  hag03}.

In what concerns the quiet Sun, much has been advanced in the last
decade thanks to the use of full Stokes polarimetry at high spatial
resolution, as provided by ground-based and space-borne telescopes
\citep[see][]{moreno:2012}.  Early spectropolarimetric measurements
from the ground revealed that also in the quiet Sun magnetic flux is
emerging continually \citep[e.g.,][]{lites:1996, bart02,marian:2007}.
The Solar Optical Telescope on board the Hinode satellite
\citep{kos07} significantly extended our observational capabilities,
making it possible to characterize granular-sized flux emergence
events with a resolution and sensitivity never reached before
\citep[e.g.,][]{cen07,ot07,os08, mg09,wang:2012}.  Using these data,
\citet{ishi08} and \citet{ishi09} demonstrated that the emergence of
magnetic flux on granular scales brings large amounts of horizontal
fields to the quiet photosphere, while \citet{luis12} proved that the
solar internetwork is indeed full of weak horizontal magnetic fields.

To characterize flux emergence events {\bf in the photosphere
  and above} from the beginning to the end, one needs simultaneous
measurements in all possible wavelength ranges, from the infrared to
hard X-rays, at the highest spatial resolution. Therefore, having a
complete observational picture of flux emergence has proven extremely
difficult.

Concerning new modeling capabilities, our knowledge has also advanced
thanks to the availability of very extensive (sunspot-sized simulation
boxes), very high-resolution, realistic 3D numerical simulations of
flux emergence in the solar atmosphere. According to \citet{tor09},
two classes of models can be distinguished.  The first type include
models with boundaries from the top of the convection zone to the
corona.  In order to boost computing speed, they strongly simplify the
thermodynamics of the plasma and ignore its interaction with the
radiation field; in most cases, heat conduction is also neglected.
Such simplifications limit the possibility of comparison with
observations in the regions where radiative transfer processes are
important \citep[see][and references therein]{tor09}. A second class
of models have appeared recently that solve the radiation transfer
problem simultaneously with the MHD equations \citep{mark07,mark08,
  juan08,tor09}. These authors injected a horizontal magnetic flux
tube or sheet (with different levels of field line twist) in the lower
levels of the domain.  \citet{juan08} used a larger box in the
vertical direction reaching up to the chromosphere, transition region
and part of the corona.  They found cold bubbles developing in
coincidence with the arrival of the magnetized plasma to the
chromosphere.  \citet{tor09} also found irregular and extended cool
patches when the magnetized plasma reaches the chromosphere. They
concluded that "the rise of the magnetized plasma in the low
atmosphere proceeds in the form of a series of jumps with stops in
between (at levels mostly between 200 and 500 km), instead of in a
continuous fashion". {\bf The reason given by \citet{tor09} for the
  occurrence of these jumps and stops is the necessity of the plasma
  to pile up at a certain height and to expand sideways so the plasma
  $\beta$ decreases and buoyancy instabilities can develop again.  Once this
  occurs the magnetized plasma continues its way up.}

The interaction between emerging flux and pre-existing ambient fields
has become a hot topic of research for both observers and modelers.
When the two flux systems interact, reconnection is likely to
happen, producing brightenings and surges. Therefore, observations of
the emerging flux and its interaction with the preexisting field are
important to understand key processes of energy release and associated
phenomena in the solar atmosphere.

Signatures of energy release at different scales and heights have been
identified above emerging flux regions repeatedly \citep[for a review,
see][]{guglielmino:2012}. The work of \citet{salvo10} is a recent
example of multiwavelength, multilayer studies of flux emergence
events. Using observations in the visible, UV, EUV and soft X-rays,
they analyzed brightenings in the chromosphere, transition region, and
corona, as well as chromospheric surges, associated with an
intermediate-scale emerging flux region and ascribed them to the
reconnection of the newly emerged flux and the preexisting field.
Other examples of energy release in active regions are small-scale
brightenings and transient emissions in the wings of the H$\alpha$
line, the so-called Ellerman bombs \citep{geor02,watanabe:2011}.

We have observed a new type of granular-sized flux emergence events
using high resolution spectropolarimetric measurements in the
photospheric \ion{Fe}{1} 630.2~nm lines and the chromospheric
\ion{Ca}{2} 854.2~nm line. These events differ from others reported in
the literature \citep[e.g.,][]{cen07,mg09,gom10,marian:2010,gom13} in
that they are not loop-shaped, but instead have a 3D semi-spherical
shape, resembling a {\em parachute}. Due to their similarities we name
them {\em magnetic bubbles}, as opposed to magnetic loops. If magnetic
loops intersect the photosphere at two points, called footpoints, our
magnetic bubbles do so in {\em half moon} shaped magnetic feet or legs
of opposite polarity. In \ion{Ca}{2}~854.2~nm images the bubbles
appear dark, therefore we simply refer to them as {\em dark bubbles}.

This paper presents a detailed characterization of these magnetic
bubbles, including their temporal evolution, their magnetic and
dynamic properties in different layers of the atmosphere, and a
comparison with numerical simulations that show similar features.
Paper II of the series will be devoted to an in-depth analysis of the
observed \ion{Ca}{2} 854.2~nm Stokes profiles, which will be inverted
and compared with synthetic spectra calculated from the simulations.

We start by describing the observations and the data reduction in
Section~\ref{data}. In Section \ref{results} we present two examples
of emerging magnetic bubbles and the various phenomena associated with
them. We also determine the velocity and magnetic structure of the
bubbles from a bisector analysis and Stokes inversion of the observed
polarization signals.  Section \ref{mhd} describes our numerical
simulations and the bubbles they produce.  Finally, in
Sect.~\ref{disc} we discuss our findings and compare them with earlier
observational results and simulations.

\section{Observations and data analysis}
\label{data}

\subsection{Observations}
\label{obs}

We analyze observations of AR 11024 acquired on 2009 July 5 using the
Swedish 1-m Solar Telescope \citep[SST;][]{2003scharmer} and the CRisp
Imaging Spectro-Polarimeter \citep[CRISP;][]{2008scharmer}. The seeing
conditions were very favorable and from the large volume of recorded
data we select a period of 28 minutes starting at 09:48:40~UT. During
this time the FOV was centered near S$27 ^\circ$, W$12 ^\circ$. The
observations consist of simultaneous full-Stokes measurements of the
\ion{Fe}{1}~630.15 and 630.25~nm lines and the \ion{Ca}{2}~854.2~nm
line. The spatial resolution of the observations is close to the
diffraction limit of the telescope ($\lambda/D=0.14\arcsec$ at 630~nm).

\begin{figure*}
\centering
\resizebox{\hsize}{!}{\includegraphics{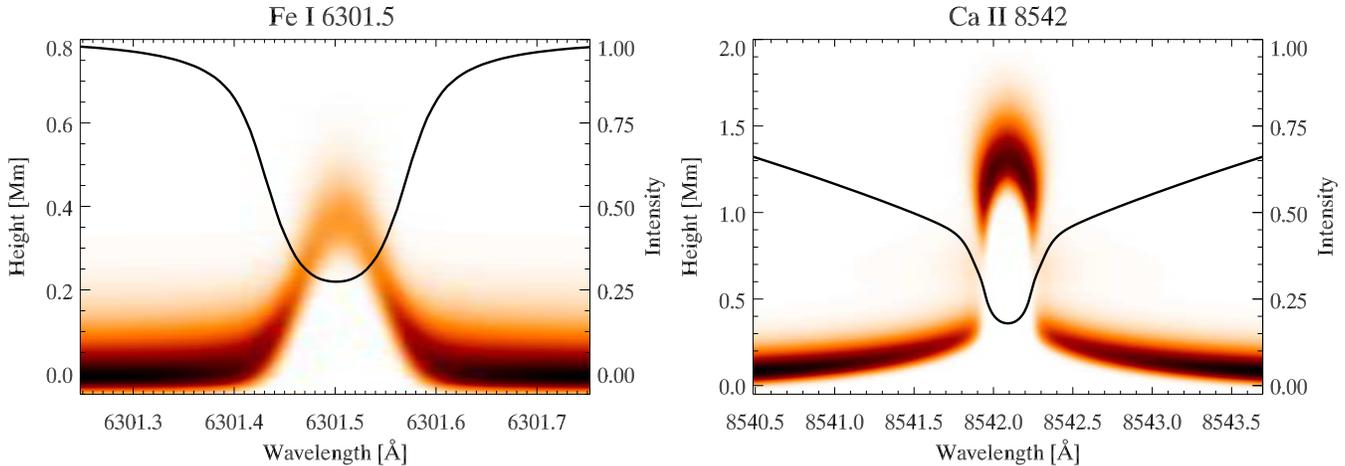}}
\caption{\emph{Left:} contribution function of the
  \ion{Fe}{1}~630.15~nm line in the FALC model. \emph{Right:}
  contribution function of the \ion{Ca}{2}~854.2~nm line. The
  corresponding intensity profiles are shown in black. Note the
  different vertical scales for each line.  The transition between
  photosphere and chromosphere occurs around 500~km and is signaled by
  the knees of \ion{Ca}{2}~854.2~nm at $\pm 300$~m\AA\/.}
\label{figcont}
\end{figure*}

CRISP is a tunable filter consisting of two etalons in telecentric
mount. This type of instruments allows one to observe the entire field
of view (FOV) in quasi-monochromatic light (the CRISP passband at
630~nm has a full width at half maximum of 65~m\AA).  Spectral lines
are scanned by tuning the etalons. Each of the \ion{Fe}{1} lines was
sampled at 15 positions across the range $\pm336$~m\AA \ from line
center in steps of 48~m\AA. A continuum point was observed at
$+680$~m\AA. The \ion{Ca}{2}~854.2~nm line was sampled at 17 positions
across the range $\pm 800$~m\AA\/ in steps of 100~m\AA, plus an extra
point located at $+2.4$~\AA.

The polarization modulation was achieved with the help of two liquid
crystal variable retarders cycling through four states. The
polarization analysis was performed by a polarizing beam splitter in
front of the two narrow-band cameras of CRISP. In order to reduce the
noise, we recorded nine exposures per modulation state, resulting in
an effective integration time of 153 ms per wavelength position and
Stokes parameter. A full spectral scan of the two \ion{Fe}{1} lines,
the \ion{Ca}{2} line and the two continuum positions was completed in
61~s. The time sequence contains 28 line scans with a cadence of 67~s.

Here we will often refer to events observed in the wings and cores of
the iron and calcium lines. To establish a rough relationship between
the various wavelength positions and geometrical heights, in
Figure~\ref{figcont} we plot the contribution functions of
\ion{Fe}{1}~630.15~nm and \ion{Ca}{2}~854.2~nm as computed in the FALC
model of \cite{1993ApJ...406..319F}. The $\tau=1$ layer is very
corrugated in the chromosphere, especially in the presence of magnetic
fields. The FALC model cannot reflect this corrugation because it is
one dimensional, but serves to illustrate the range of heights that
we can expect for \ion{Ca}{2} 854.2~nm. Figure~\ref{figcont} shows
that all phenomena observed in the \ion{Fe}{1} lines happen within the
photosphere (the wings referring to the low photosphere close to the
continuum and the core to the upper photosphere). {\bf Events detected
  in the wings of the \ion{Ca}{2}~854.2~nm line take place in the
  mid-high photosphere, while events observed in the line core occur
  in the mid chromosphere}. For example, the \ion{Ca}{2} filtergrams
at $-0.06$~nm from the line core probe the photosphere at a height of
about 200~km, and hence show reversed granulation. The ``knees'' of
the intensity profile at around $\pm 300$~m\AA\/ sample the minimum
temperature region, some 500~km above the continuum forming layer. 
The line core, representing heights of approximately 1300~km, is 
purely chromospheric.

\begin{figure*}
\centering
\resizebox{\hsize}{!}{\includegraphics{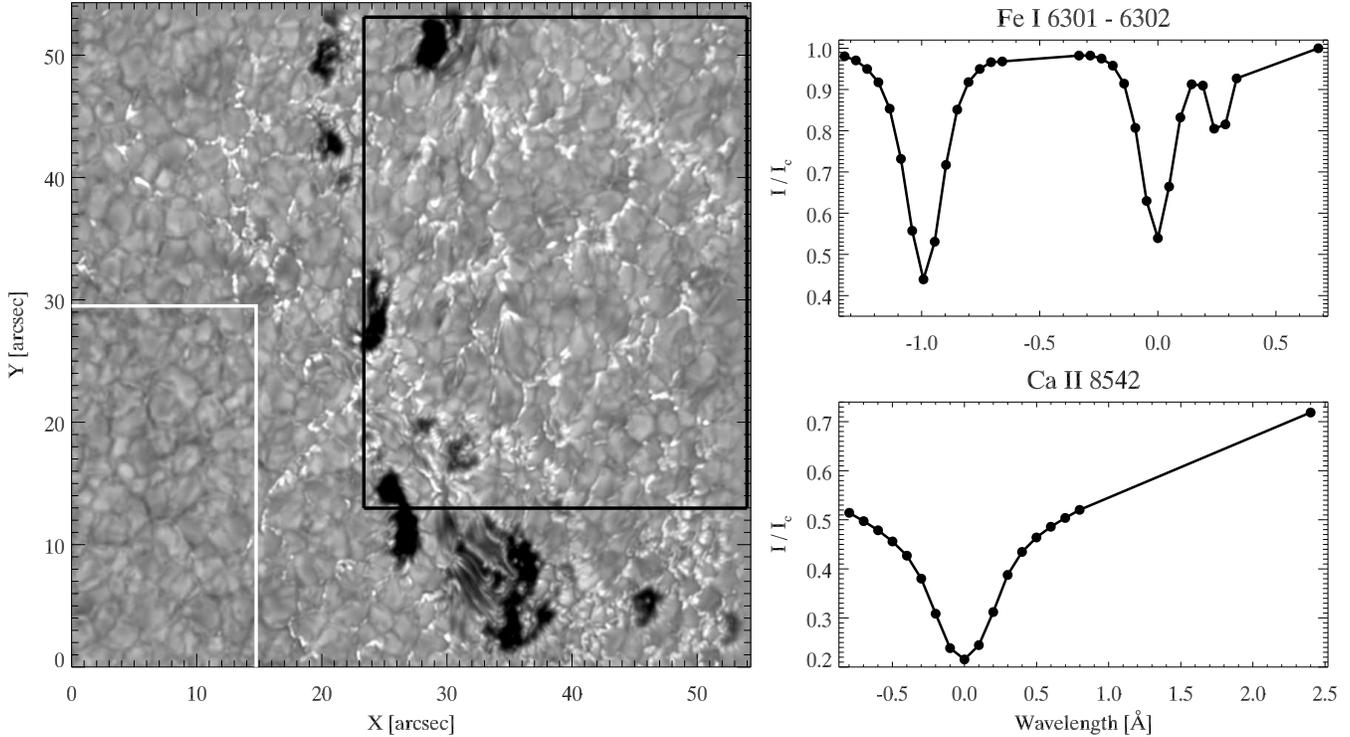}}
\caption{\emph{Left:} Part of active region NOAA 11024 observed $+
  2.4$~\AA \ from the \ion{Ca}{2}~854.2~nm line center. The black
  rectangle on the right highligths the FOV chosen for
  Figure~\ref{fig1}.  \emph{Right:} Spatially-averaged quiet-Sun
  profiles of the \ion{Fe}{1}~630.15 and 630.25~nm ({\rm top}) and
  \ion{Ca}{2}~854.2~nm lines ({\rm bottom}), computed within the white
  box shown on the left panel. The observations were acquired at the
  line positions marked with circles.  \vspace*{1em}}
\label{fig0}
\end{figure*}

Through the use of the SST adaptive optics system and the
multi-object, multi-frame blind deconvolution image restoration
technique \citep[MOMFBD;][]{2005SoPh..228..191V}, we correct for most
of the wavefront distortions and image blurring induced by the Earth's
atmosphere. Precise alignment between the wideband and narrowband
cameras is achieved by a separate alignment procedure involving a
reference pinhole array target. The pre-processing and polarimetric
calibration of the data follow the methods described in
\citet{jaime:2013}, and telescope-induced polarization was
compensated at 630~nm and 854~nm using the telescope models of
\citet{2010selbing} and \citet{2010delacruz}, respectively. For each
spectral line, the entire time series was aligned and de-rotated to a
common reference, and residual rubbersheet motions were removed as
described by \citet{1994shine}. Finally, the \ion{Ca}{2}~854.2~nm
images in each scan were aligned with respect to the
\ion{Fe}{1}~630.15 images from the same time step. To achieve high
accuracy, the alignment was done using the wideband images because 
they show photospheric scenes in the two spectral ranges. 

An inspection of the resulting Stokes profiles revealed a small amount
of residual crosstalk from Stokes I into Q, U, and V.  We removed this
contamination as follows,
\begin{equation}
  S(\lambda) = S_{\rm obs}(\lambda) - S_{\rm offset} \cdot I_{\rm obs}(\lambda),
\end{equation}
where $S$ represents any of the three Stokes parameters Q, U, and V,
and $S_{\rm offset}$ is the corresponding crosstalk coefficient. The
various $S_{\rm offset}$ were calculated by linear regression of
Q($\lambda$), U($\lambda$), and V($\lambda$) against I($\lambda$), for
pixels with negligible polarization signals, and turn out to be of the
order of 0.2\%.  Finally, we removed small spectral gradients observed 
in the Stokes profiles.  After calibration, the noise in Stokes Q, U, and
V is $2.5 \times 10^{-3}$, $1.5 \times 10^{-3}$, and $1.5 \times
10^{-3}$, respectively.

The observed FOV is shown in Fig.~\ref{fig0} as seen in the red wing
of the \ion{Ca}{2}~854.2~nm line. Magnetic bright points are prominent
at this wavelength because the granulation contrast is lower than in
the continuum. A train of pores (concentrations of magnetic field
without a penumbra) appears embedded in the network of bright points.
The observed line positions are indicated with circles in the right
panel of the figure, superposed on spatially-averaged quiet Sun
profiles computed within the white box marked in the left panel.

\begin{figure*}
\centering
\resizebox{\hsize}{!}{\includegraphics{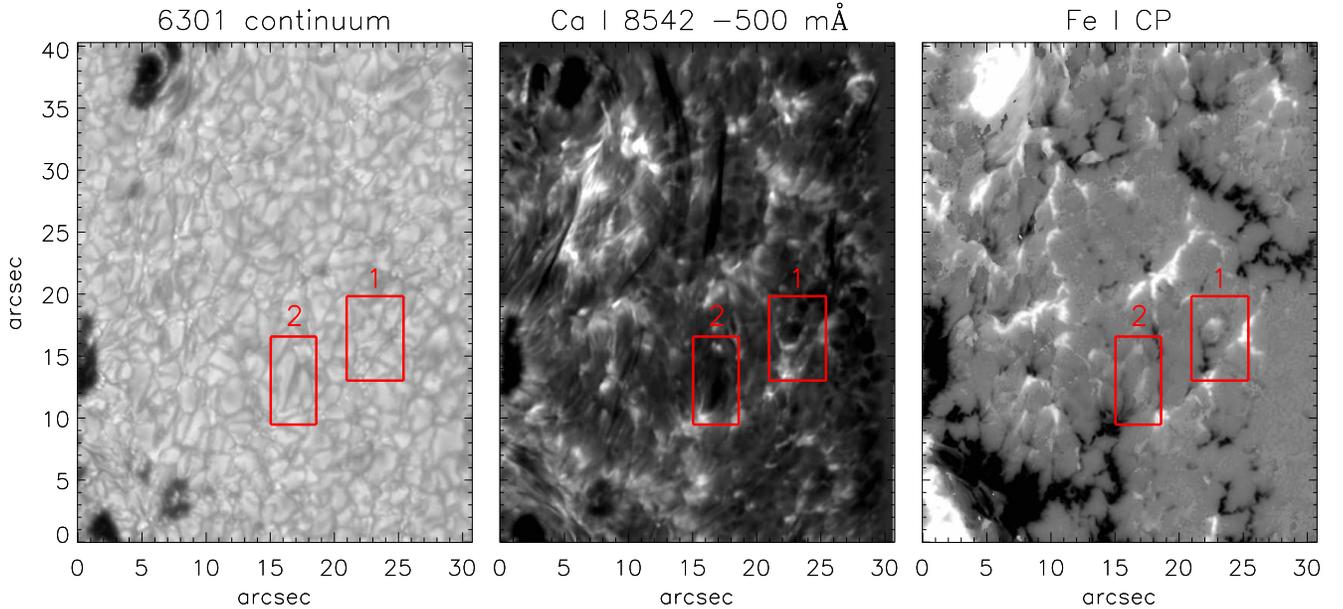}}
\caption{SST observations of NOAA 11024 on 2009 July 5 at 10:13:14~UT.
  The $30\arcsec \times 40\arcsec$ FOV shown here focuses on the area
  where magnetic flux was emerging more intensively.  {\em Left:}
  Continuum intensity at 630.32~nm.  {\em Middle:}
  \ion{Ca}{2} 854.2~nm Stokes I filtergram at $-0.05$~nm from line center.
  {\em Right:} mean circular polarization degree in the \ion{Fe}{1}
  lines (scaled to $\pm$5\%).  The red boxes outline the FOVs
corresponding to each of the two emergence events analyzed in this
paper (cases \#1 and \#2).
\label{fig1}}
\end{figure*}

AR 11024 appeared on 2009 July 4 as a pore that developed a penumbra
on the same day \citep{schliche:2010, schliche:2012}. Flux emergence
was taking place intensively everywhere in the AR, yielding numerous
pores and sunspots. Here we focus on the quietest part of the FOV,
which also showed many instances of flux emergence on small spatial
scales. The two examples considered in this paper (cases \#1 and \#2)
are highlighted with white boxes in the intensity and circular
polarization maps of Figure~\ref{fig1}.

\subsection{Line parameters}
\label{lin}

We have calculated Dopper velocities at different heights in the
atmosphere from the bisectors of the \ion{Fe}{1}~630.15~nm line at ten
intensity levels, from 0\% (line core) to 90\% (line wing near the
continuum). To this end, we used the bisec\_abs\_lin.pro routine of
the Kiepenheuer-Institut f\"ur Sonnenphysik IDL library.  For the
calculation of bisectors, the intensity profile is interpolated
linearly. The wavelength position of the line core was obtained by
means of a parabolic fit around the intensity minimum.

In order to convert bisector positions into Doppler velocities we
assume that, on average, quiet Sun areas within the FOV are at rest.
These areas are defined as those pixels whose magnetic signals are
smaller than 3$\sigma$, with $\sigma$ the noise of the photospheric
magnetograms. Then we averaged the bisectors between the 40\% and 70\%
intensity levels in the selected pixels, and took the result as the
zero point for the velocity scale.

A similar procedure was applied to the chromospheric line. In order to
convert the \ion{Ca}{2} 854.2~nm bisector positions to velocities we
again assumed that the quiet Sun pixels are at rest on average.

To study the evolution of the longitudinal and transverse components
of the vector magnetic field we use the mean linear polarization
degree (LP),
\begin{equation}
{\rm LP} = \frac{\int^{\lambda _{\rm r}}_{\lambda _{\rm b}} 
[Q^2(\lambda) + U^2(\lambda)]^{1/2}/I(\lambda) \, 
{\rm d} \lambda}{\int^{\lambda _{r}}_{\lambda _{b}} {\rm d} \lambda},
\end{equation}
and the mean circular polarization degree (CP),
\begin{equation} {\rm CP} = \frac{ \int^{\lambda _{\rm r}}_{\lambda
      _{\rm b}} |V(\lambda)|/I(\lambda) \, {\rm d}
    \lambda}{\int^{\lambda _{r}}_{\lambda _{b}} {\rm d} \lambda},
\end{equation}
where $\lambda _{\rm r}$ and $\lambda_{\rm b}$ represent the limits of
integration over the line. The LP and CP have been calculated for the
two \ion{Fe}{1} lines separately and then averaged, in order to
improve the signal-to-noise ratio.

\subsection{Retrieval of the magnetic properties of the bubble}
\label{mag}

The four Stokes profiles of the \ion{Fe}{1} lines were inverted
simultaneously to determine the magnetic and dynamic properties of the
fields emerging in the photosphere. To this end, we used the SIR
inversion code \citep[Stokes Inversion based on Response
functions;][]{sir} and a simple one-component model atmosphere.  The
atmospheric parameters were taken to be constant with height except
for the temperature, whose stratification was obtained by perturbing
the Harvard Smithsonian Reference Atmosphere \citep{gingerich} at two
specific optical depths called nodes. In view of the high spatial
resolution attained by our observations, we set the magnetic filling
factor to unity and did not use any macroturbulence or stray light
contamination. In total, the inversion returned the values of 7 free
parameters (the three components of the vector magnetic field, the LOS
velocity, the temperature at the two nodes, and the microturbulence)
for each pixel within the observed FOV.

The longitudinal component of the chromospheric magnetic field was
computed from the \ion{Ca}{2} 854.2~nm Stokes V profiles using the
weak field approximation. We can apply this approximation here because
the Zeeman splitting is much smaller than the thermal width of the
line. The magnetic field may change over the line formation region,
though, so the method remains qualitative. In the weak field regime,
$V(\lambda)$ is related to the longitudinal component of the field
through the following relationship 
\begin{equation}
\label{wap}
{V(\lambda)} = -\phi\,C\, \frac{\delta I}{\delta \lambda},
\end{equation}
\citep[e.g.,][]{landi04}, where $\phi=f\, B\, \cos(\gamma)$ is the
longitudinal flux density, $f$ the magnetic filling factor, $B$ the
field strength, $\gamma$ the field inclination with respect to the
LOS, $C=4.67\times 10^{-13}\lambda_{0}^2g$ a constant that depends on
the central wavelength $\lambda_{0}$ and the effective Land\'e factor
$g$ of the transition, and $\delta I/{\delta \lambda}$ the derivative
of the intensity profile with wavelength. $\phi$ is in Mx~cm$^{-2}$
when $\lambda_{0}$ is given in~\AA. We have followed the least-squares
method used by \citet{mg09} to obtain $\phi$ as
\begin{equation}
\label{campo}
\phi = - \frac{\Sigma_{i} \frac{\delta I}{\delta \lambda_{i}}V_{i}}{C\,
\Sigma_{i}(\frac{\delta I}{\delta \lambda_{i}})^2},
\end{equation}
where the subscript $i$ labels the $i^{\rm th}$ wavelength sample.

\section{Results}
\label{results}

In this Section we describe two characteristic flux emergence events
and the various phenomena that occur simultaneously as the magnetic
field crosses the photosphere and intrudes at least into the mid
chromosphere.





{\bf After a global description of the two emergence events based on
  the recorded filtergrams, we will perform a more detailed analysis
  of event~\#1 that will include a discussion of the properties of the
  observed Stokes profiles and a derivation of LOS velocities and
  magnetic fields both in the photosphere and in the chromosphere. }

\subsection{The dark bubble and other signs of flux emergence}

\begin{figure*}
\centering
\resizebox{0.7\hsize}{!}{\includegraphics{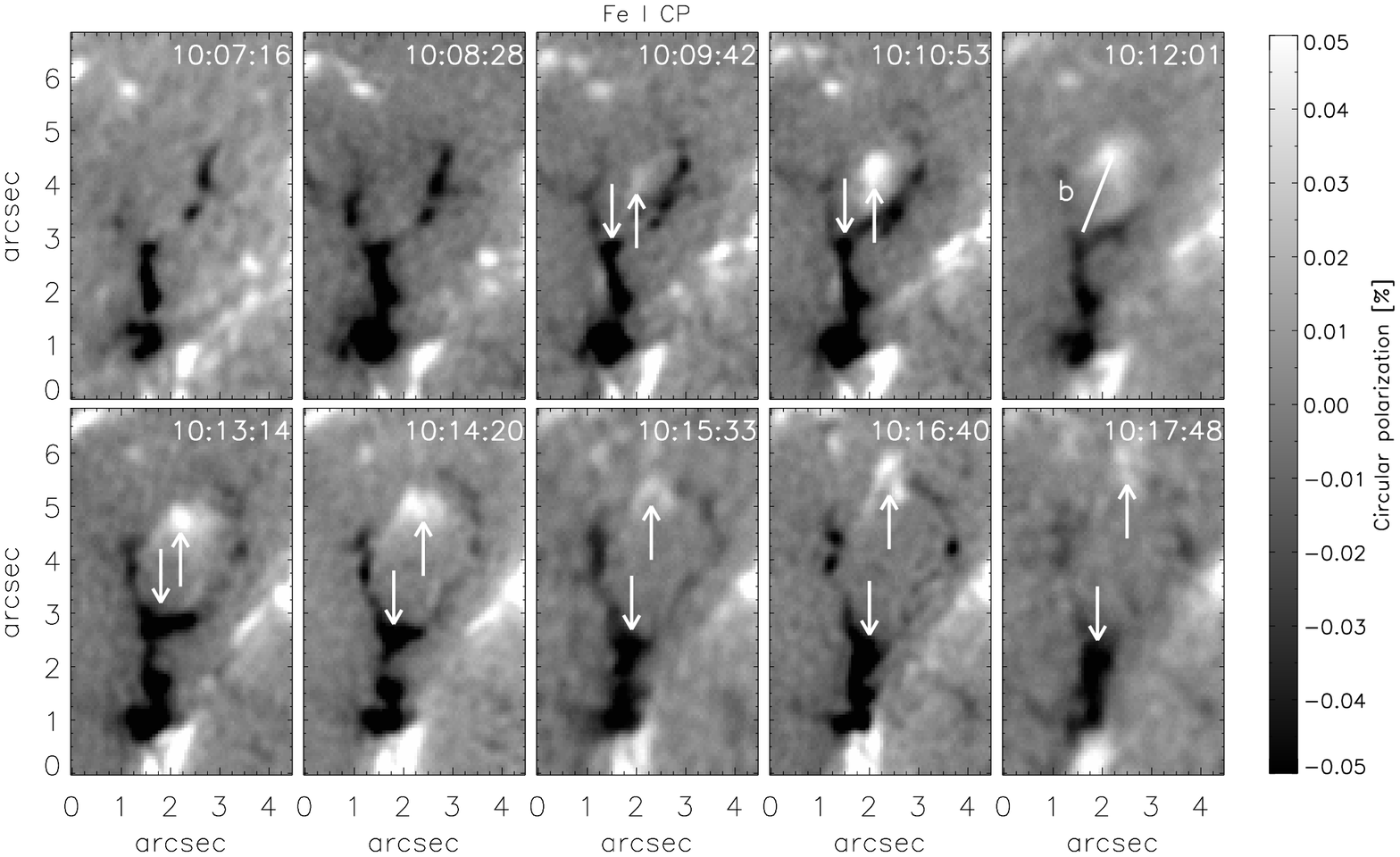}} \\
\resizebox{0.7\hsize}{!}{\includegraphics{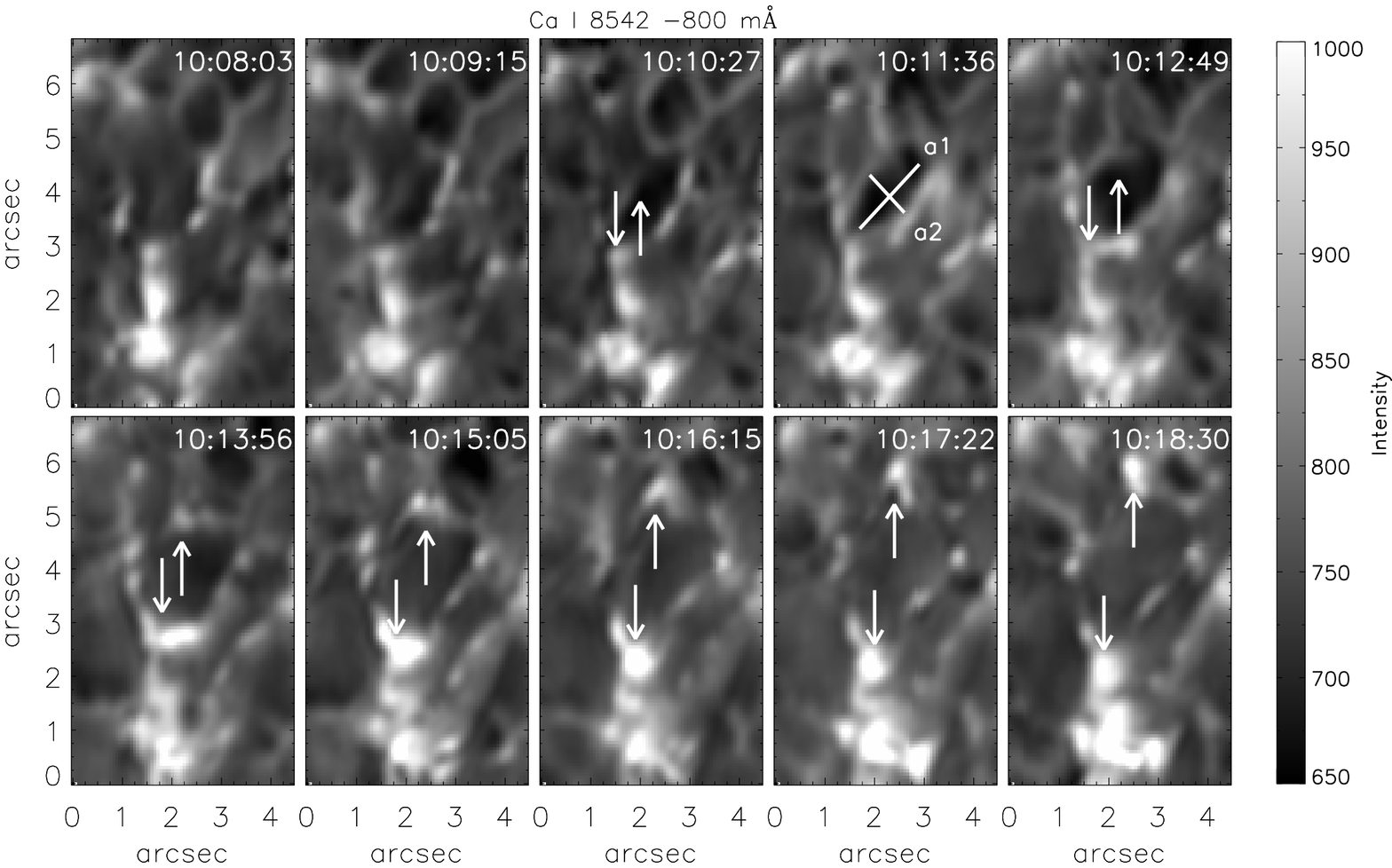}}

\caption{{\em Top:} Temporal evolution of the photospheric
  magnetograms corresponding to the first case of flux emergence. Time
  in~UT is indicated in the upper right corner of each panel. The
  sequence shows the appearance of a positive polarity foot (white)
  close to an already existing negative polarity semi-circular patch
  (black). The arrows indicate feet of opposite polarity as they
  separate from each other during the emergence process. {\em Bottom:}
  same as above for the \ion{Ca}{2} 854.2 $-0.08$~nm intensity
  filtergrams.  Here the arrows still highlight the magnetic legs.
  Major intensity enhancements are associated with the negative leg.
  Some of the brightenings [located around
  $(x,y)=(2\arcsec,0.5\arcsec)$] appear when the negative foot interacts
  with a pre-existing positive patch. The dark bubble is well visible
  in the intensity images from 10:10:27 to 10:13:56~UT. $a_1$ and
  $a_2$ are its dimensions. $b$ is the distance between the magnetic
  legs of opposite polarity (Tables ~\ref{table1} and
  ~\ref{table2}).\label{fig2}}

\end{figure*}

Figure~\ref{fig2} displays a temporal sequence of CP maps (upper
panels) and \ion{Ca}{2} 854.2 $-0.08$~nm filtergrams (lower panels)
for flux emergence event \#1. The times indicated in each panel are
those of the first wavelength point of the corresponding
\ion{Fe}{1}~630~nm or \ion{Ca}{2} 854.2~nm line scan, in~UT.

During the sequence, which spans about 10~minutes, a negative polarity
patch showing a semi-circular shape is observed. In the third
magnetogram of the sequence, a weak positive blob appears at
$(x,y)=(2\arcsec,4\arcsec)$.  This magnetic knot rapidly grows from
the third to the fourth magnetogram, reaching a size of $0.5\arcsec
\times 0.8\arcsec$. The subsequent evolution is dominated by the
separation of the opposite magnetic polarities, as indicated by the
arrows. At 10:09:42~UT the centers of the two polarities are separated
by 1.4\arcsec.  This distance grows steadily up to 3.6\arcsec at
10:17:48~UT.  The separation speed reaches a maximum of 5.0
km\,s$^{-1}$ at 10:14:20~UT and slows down to 2.7 km\,s$^{-1}$ at the
end of the sequence.

\begin{figure}
\centering
\resizebox{\hsize}{!}{\includegraphics{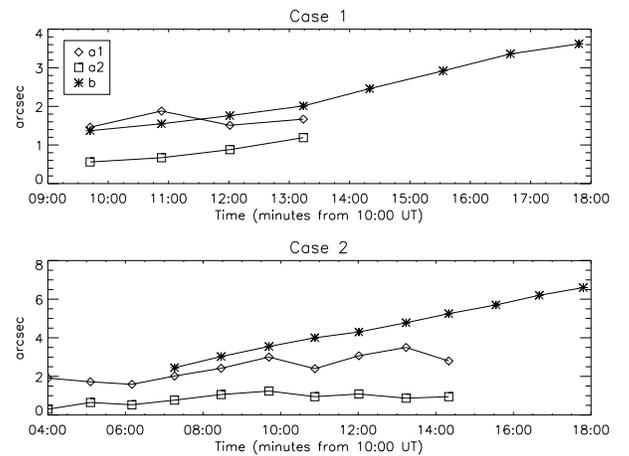}}
\caption{Temporal evolution of the $a_1$, $a_2$, and $b$ parameters
  for cases \#1 and \#2. $a_1$ and $a_2$ are the dimensions of the
  dark intensity bubble as depicted in Figure~\ref{fig2}. The bubble
  is defined as the pixels with \ion{Ca}{2} intensities equal to $0.7
  \, I_{\rm c}$ or less, that are within the area enclosed by the
  magnetic feet. $b$ is the distance between magnetic legs of opposite
  polarity, also depicted in Figure~\ref{fig2} (see
  Tables~\ref{table1} and ~\ref{table2} for numerical
  values).\label{figparavstime}}
\end{figure}

The lower panels of Figure~\ref{fig2} show the corresponding
\ion{Ca}{2} intensity images. The arrows still refer to the magnetic
legs described above. Brightenings in the intensity sequence resemble
in shape and coincide in position with the magnetic features seen in
the magnetograms. Particularly bright areas are observed from
10:13:56~UT until the end of the sequence. However, the most striking
feature in the intensity images is the occurrence of a dark bubble,
which occupies a more or less circular area within the limits defined
by the magnetic feet. For the purpose of detection, we have defined
the dark bubble as those pixels -- within the mentioned limits --
whose intensity in the \ion{Ca}{2} 854.2 $-0.08$~nm filtergram is
$\leq$ 0.7$I_{\rm c}$. Using this criterion, the bubble turns out to
be visible from 10:09:42 to 10:13:14~UT (magnetogram time), i.e., for
four line scans. The dark bubble grows from $1.5\arcsec \times 0.6\arcsec$
at 10:09:42~UT to $1.7\arcsec \times 1.2\arcsec$ at 10:13:14~UT.  The
evolution of the bubble size (the $a_1$ and $a_2$ parameters) as well as
the separation between the centroids of the opposite-polarity magnetic
legs (the $b$ parameter) is summarized in Table ~\ref{table1} and
Figure~\ref{figparavstime}.  We find a more or less steady increase in
the $a_2$ direction, but an intermittent increase in the $a_1$
direction. This is also reflected in the expansion velocity, which is
steady along the $a_2$ axis (from 1.1 to 3 km\,s$^{-1}$) but can take
both positive and negative values along the $a_1$ axis. Note, for
example, the negative expansion velocity of $-3.9$ km\,s$^{-1}$ at
10:12:01~UT.

\begin{deluxetable*}{cccccccc}
\tablecolumns{8}
\tablewidth{0pc}
\tablecaption{Properties of \ion{Ca}{2} 854.2 bubble \#1 \label{table1}}
\tablehead{
\colhead{Time} & \colhead{Bubble size} &  \colhead{Magnetic feet separation} & \multicolumn{2}{c}{Bubble horizontal
expansion} & \colhead{Magnetic feet horizontal speed} & \colhead{Vertical speed} & \colhead{Rising
time} \\ 
\colhead{(UT)} & \colhead{($a_1 \times a_2$)} & \colhead{($b$)} & \colhead{($a_1$, km\,s$^{-1}$)} & \colhead{($a_2$, km\,s$^{-1}$)}
&\colhead{(km\,s$^{-1}$)} & \colhead{(m\AA\,s$^{-1}$)} & \colhead{(s)} 
}
\startdata
10:07:16 & \nodata & \nodata & \nodata & \nodata & \nodata & \nodata & \nodata \\
10:08:28 & \nodata & \nodata & \nodata & \nodata & \nodata & \nodata & \nodata \\
10:09:42 & $1.46\arcsec \times 0.56\arcsec$ & 1.37\arcsec & \nodata & \nodata  & \nodata & \nodata & \nodata \\
10:10:53 & $1.88\arcsec \times 0.67\arcsec$ & 1.55\arcsec & 4.23 & 1.10 & 1.81 & 1.38 & 72 \\
10:12:01 & $1.51\arcsec \times 0.88\arcsec$ & 1.76\arcsec & $-3.88$ & 2.20 & 2.20 & 1.45 & 69 \\
10:13:14 & $1.67\arcsec \times 1.19\arcsec$ & 2.01\arcsec & 1.56 & 3.03 & 2.45 & 1.35 & 74 \\
10:14:20 & \nodata & 2.46\arcsec & \nodata & \nodata & 4.86 & \nodata & \nodata \\
10:15:33 & \nodata & 2.92\arcsec & \nodata & \nodata & 4.50 & \nodata & \nodata \\
10:16:40 & \nodata & 3.36\arcsec & \nodata & \nodata & 4.69 & \nodata & \nodata \\
10:17:48 & \nodata & 3.62\arcsec & \nodata & \nodata & 2.73 & \nodata & \nodata
\enddata
\tablecomments{ Column (1): Time of the first wavelength of a full
  scan (2): Size of dark bubble as measured in \ion{Ca}{2} intensity.
  The bubble is defined by the intensities of less than $0.7 \, I_{\rm
    c}$ that are within the area enclosed by the magnetic feet.
  Dimensions $a_1$ and $a_2$ are defined in Figure~\ref{fig2} and
  displayed in Figure~\ref{figparavstime} (3): Distance between the
  opposite-polariy magnetic feet. The separation $b$ is defined in
  Figure~\ref{fig2} and summarized in Figure~\ref{figparavstime} (4):
  Horizontal expansion of the bubble area in the $a_1$ and $a_2$
  directions relative to the previous scan (5): Speed of separation of
  the opposite-polarity magnetic feet (6): Vertical speed of the
  bubble measured from one wavelength position (or height) to the next
  (6): Bubble rising time from one wavelength position (or height) to
  the next. {\bf Missing information is marked with the symbol ``\nodata''.}}
\end{deluxetable*}

\begin{deluxetable*}{cccccccc}
\tablecolumns{8}
\tablewidth{0pc}
\tablecaption{Properties of \ion{Ca}{2} 854.2 bubble \#2 \label{table2}}
\tablehead{
\colhead{Time} & \colhead{Bubble size} &  \colhead{Magnetic feet separation} & \multicolumn{2}{c}{Bubble horizontal
expansion} & \colhead{Magnetic feet horizontal speed} & \colhead{Vertical speed} & \colhead{Rising
time} \\ 
\colhead{(UT)} & \colhead{($a_1 \times a_2$)} & \colhead{($b$)} & \colhead{($a_1$, km\,s$^{-1}$)} & \colhead{($a_2$,
km\,s$^{-1}$)}
&\colhead{(km\,s$^{-1}$)} & \colhead{(m\AA\,s$^{-1}$)} & \colhead{(s)} 
}
\startdata
10:04:00 & $1.91\arcsec \times 0.30\arcsec$ & \nodata & \nodata & \nodata & \nodata & \nodata & \nodata \\
10:05:06 & $1.72\arcsec \times 0.65\arcsec$ & \nodata & -2.00 & 3.78 & \nodata & \nodata & \nodata \\
10:06:10 & $1.59\arcsec \times 0.53\arcsec$ & \nodata & -1.45 & -1.33 & \nodata & 1.53 & 65 \\
10:07:16 & $2.02\arcsec \times 0.77\arcsec$ & 2.45\arcsec & 4.65 & 2.59 & \nodata & 1.50 & 67 \\
10:08:28 & $2.42\arcsec \times 1.06\arcsec$ & 3.03\arcsec & 3.97 & 2.88 & 5.76 & \nodata & \nodata \\
10:09:42 & $3.00\arcsec \times 1.24\arcsec$ & 3.55\arcsec & 5.60 & 1.74  & 5.02 & \nodata & \nodata \\
10:10:53 & $2.40\arcsec \times 0.95\arcsec$ & 4.00\arcsec & -6.00 & -3.00 & 4.53 & \nodata & \nodata \\
10:12:01 & $3.07\arcsec \times 1.09\arcsec$ & 4.30\arcsec & 7.00 & 1.50 & 3.15 & \nodata & \nodata \\
10:13:14 & $3.50\arcsec \times 0.87\arcsec$ & 4.78\arcsec & 4.21 & -2.15 & 4.70 & \nodata & \nodata \\
10:14:20 & $2.80\arcsec \times 0.95\arcsec$ & 5.25\arcsec & -7.57 & 0.86 & 5.08 & \nodata & \nodata \\
10:15:33 & \nodata & 5.70\arcsec & \nodata & \nodata & 4.40 & \nodata & \nodata \\
10:16:40 & \nodata & 6.20\arcsec & \nodata & \nodata & 5.30 & \nodata & \nodata \\
10:17:48 & \nodata & 6.60\arcsec & \nodata & \nodata & 4.20 & \nodata & \nodata
\enddata
\tablecomments{
{\bf Same as Table~\ref{table1}, for flux emergence event \#2}. \vspace*{.3cm} }  
\end{deluxetable*}

\begin{figure*}
\centering
\resizebox{0.8\hsize}{!}{\includegraphics{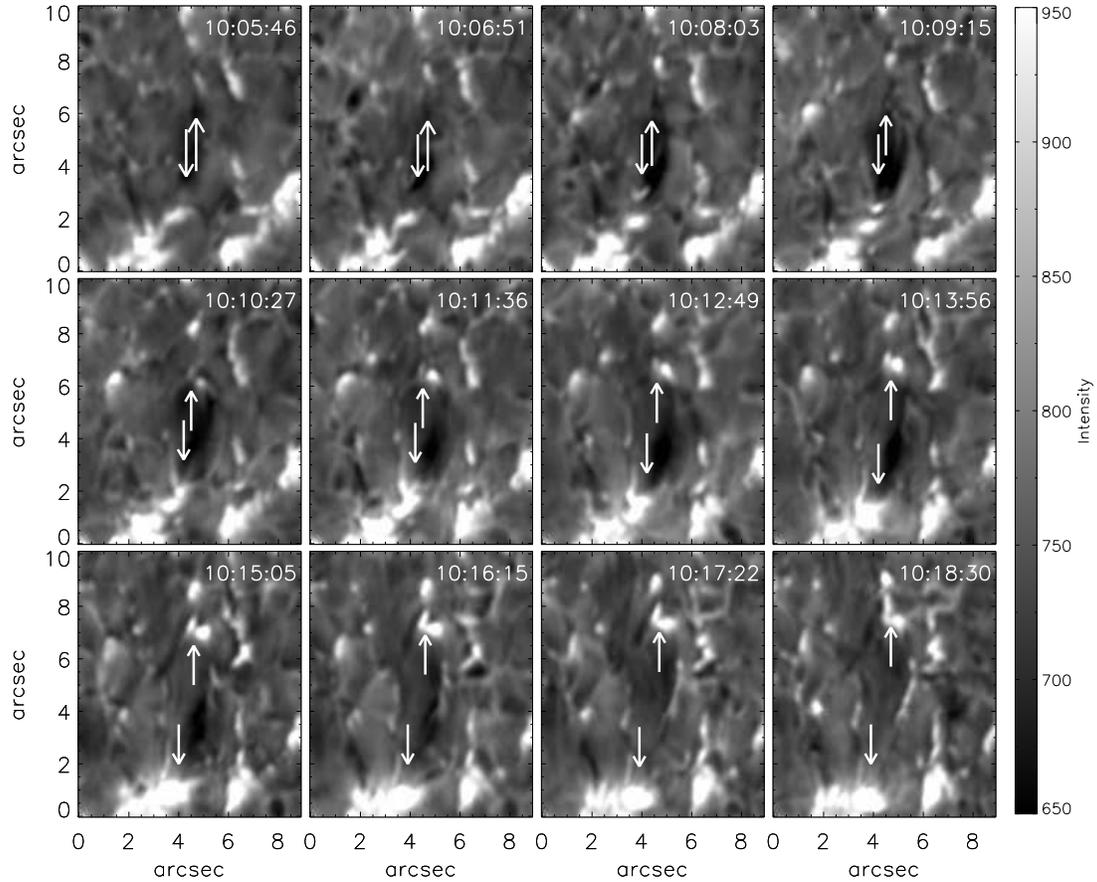}}
\caption{Same as Figure~\ref{fig2}, for the second example of flux
  emergence. In this case, only the \ion{Ca}{2} 854.2~nm intensity
  filtergrams are shown. The arrows pinpoint the opposite-polarity
  magnetic feet as they separate from each other during the emergence
  process. Note the brightenings at $(x,y)=(1\arcsec-4\arcsec,
  0\arcsec-2\arcsec)$. In this event, the dark bubble is more
  elongated and lasts during the whole 12 min sequence. \label{fig3}}
\end{figure*}

The evolution of flux emergence event~\#2 is shown in
Figure~\ref{fig3}. Only the \ion{Ca}{2} 854.2~nm intensity
filtergrams are displayed, but the details of the magnetic feet
separation and the bubble growth can be found in Table~\ref{table2}
and Figure~\ref{figparavstime}. The sequence lasts for 12~minutes and
the general behavior is very similar to case \#1.  The arrows pinpoint
the magnetic feet of opposite polarity as they separate from each
other during the emergence process. In this case, the feet reach a
maximum separation of 6.6\arcsec after 10 minutes.  The dark bubble
grows in size, but in a rather discontinuous way, suggesting that the
emergence of flux occurs in the form of successive jumps with
intermediate stops. {\bf The two events considered here show the same
  behavior}. The horizontal expansion velocity of the dark bubble
peaks at 7~km\,s$^{-1}$, with an average of 1~km\,s$^{-1}$.

\begin{figure*}
\centering
\resizebox{0.8\hsize}{!}{\includegraphics{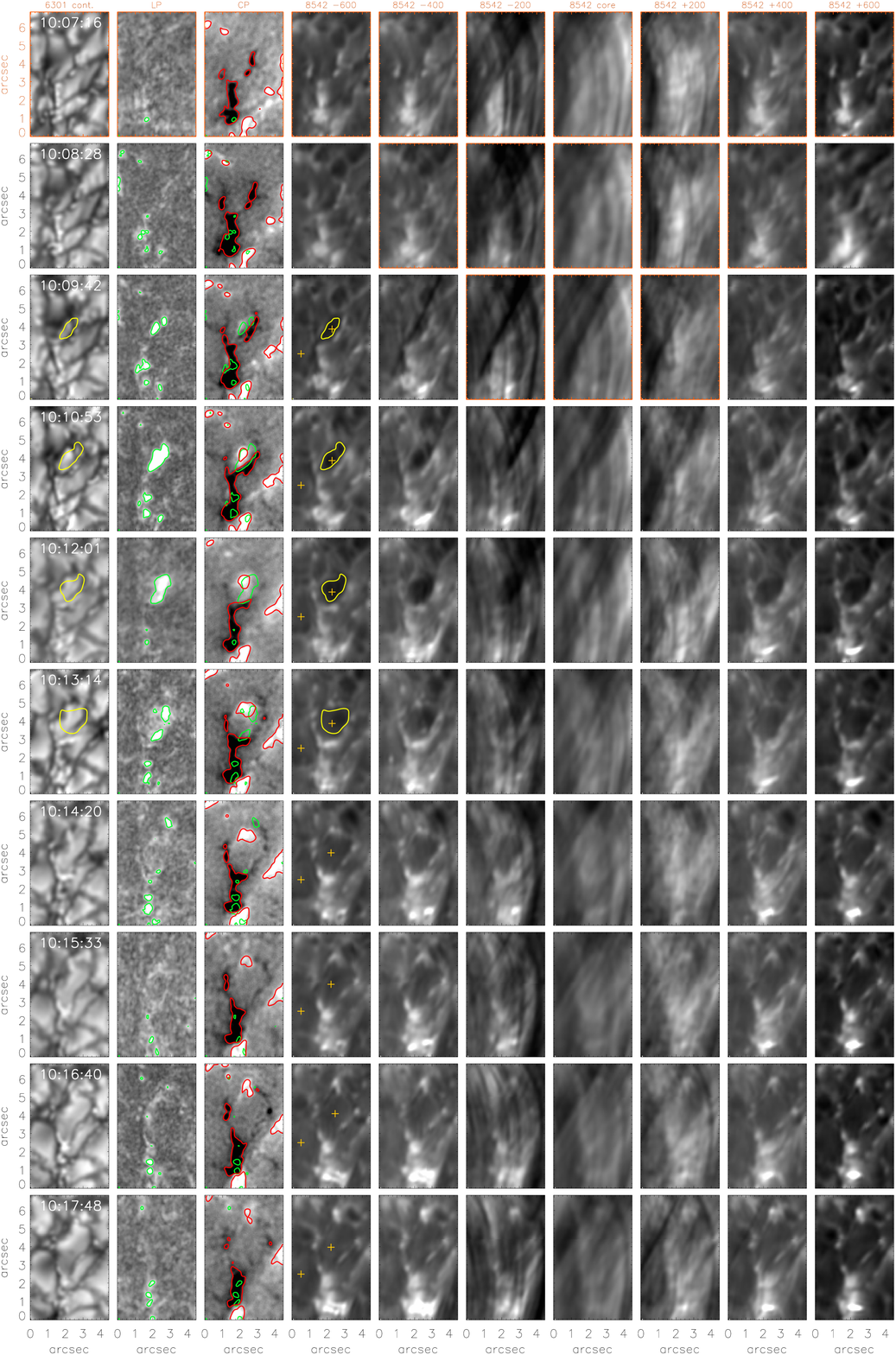}}
\vspace{0.35cm}
\caption{Temporal evolution of several observables corresponding to
  flux emergence case \#1.  From left to right: continuum intensity 
  at 630~nm, photospheric LP and CP maps, and filtergrams across 
  the \ion{Ca}{2} 854.2~nm line from $-0.06$ to $+0.06$~nm in
  steps of 0.02~nm. The green contours overplotted on the LP and CP 
  panels indicate linear polarization signals of 0.6\% of the $I_{\rm c}$. 
  The red contours shown on the CP maps represent circular polarization 
  signals of 1.3\%. The yellow contours mark the presence of 
  the dark bubble during its life. Note that the dark bubble appears 
  only in four line scans, while the magnetic bubble lives longer. 
  Contours have been plotted only in two columns (1 and 4) to avoid
  overcrowding. Pixels marked with a yellow cross are those used 
  in Figures~\ref{fig10} and \ref{fig9}. The rise of the bubble through 
  the atmosphere can be followed as it progresses from the wings to 
  the core of \ion{Ca}{2} 854.2~nm. Panels with an orange frame are 
  those in which the dark bubble is not yet visible. 
  \label{fig4}}
\end{figure*}

To get a better impression of how these events evolve with height, Figure~\ref{fig4} shows a temporal sequence of
observables formed at various heights for flux emergence case \#1 {\bf (see also the accompanying movie in the
electronic edition of the Journal)}. The sequence spans 10 minutes.  From left to right the following parameters are
displayed: continuum intensity at 630.32~nm, photospheric LP and CP maps, and seven filtergrams across the \ion{Ca}{2}
854.2~nm line from $-0.06$ to $+0.06$~nm in steps of 0.02~nm.  Figure~\ref{figcont} provides an approximate
translation of these line positions into heights.  The green contours in the LP and CP panels indicate a mean linear
polarization degree of 0.6\%.  The red contours shown on the CP map represent mean circular polarization degrees of
1.3\%. The yellow contours highlight the presence of the dark bubble during its lifetime. They are plotted only in
columns 1 and 4 to avoid cluttering.

Figure~\ref{fig4} illustrates the various phenomena occurring as the
magnetic flux rises through the solar atmosphere. The most remarkable
is the formation of a dark bubble. The continuum intensity at 630.2~nm
does not show any traces of the dark bubble, as can be seen in the
first column.  However, other effects of the emergence are evident in
the photosphere, such as abnormal granulation. At the position of the
emerging flux the granules become bigger and more elongated than
usual, e.g.\ near $(x,y)=(1.5\arcsec,4.5\arcsec)$ at 10:07:16~UT.

The dark bubble rises through the solar atmosphere. This is clearly
inferred from our sampling of the \ion{Ca}{2} 854.2~nm line because we
observe the bubble to gradually move to the line center as time
progresses. The dark bubble first appears at 10:09:42~UT in the
\ion{Ca}{2} 854.2$-0.06$ and $-0.04$~nm filtergrams, that is, the mid
photospheric wings of the line. However, in the next scan at
10:10:53~UT, it is also detected in the \ion{Ca}{2} 854.2 $-0.02$~nm
filtergram and has increased in size. Subsequently, the dark bubble
grows (see Table~\ref{table1}) and shows up closer and closer to the
line core. The pattern is similar in the red wing of the line.

The third column of Figure~\ref{fig4} shows patches of enhanced
circular polarization (red contours) from the beginning of the
sequence in spatial coincidence with the magnetic feet of negative
polarity.  At 10:09:42~UT a small patch of linear polarization appears
right next to where the positive polarity patch is emerging (see upper
panels of Figure~\ref{fig2} for a more detailed view). The LP patch
has grown at 10:10:53~UT and is now in between the opposite polarities
of the bipole. This implies the rising of a horizontal magnetic field
in photospheric layers, with more vertical legs anchored deeper down.
At 10:12:01~UT the LP patch has decreased in size, but still remains
between the opposite-polarity feet. At 10:13:14~UT the linear
polarization signal is almost washed out, whereas the vertical field
is visible until the end of the sequence. The dark bubble mimics the
behavior of the LP patch: it emerges, increases in size, and lasts for
the same duration, disappearing at 10:13:14~UT. The dark bubble
criterion is only satisfied for four time scans, from 10:09:42 UT
until 10:13:14 UT. Beyond 10:13:14 UT we can still see ring-shaped
brightenings that coincide with the magnetic legs. Those are the
boundaries of the magnetic bubble. Overall, the observations indicate
that the dark bubble is a counterpart, in higher layers, of the
appearance of horizontal fields in the photosphere.

\begin{figure*}
\centering
\resizebox{0.95\hsize}{!}{\includegraphics{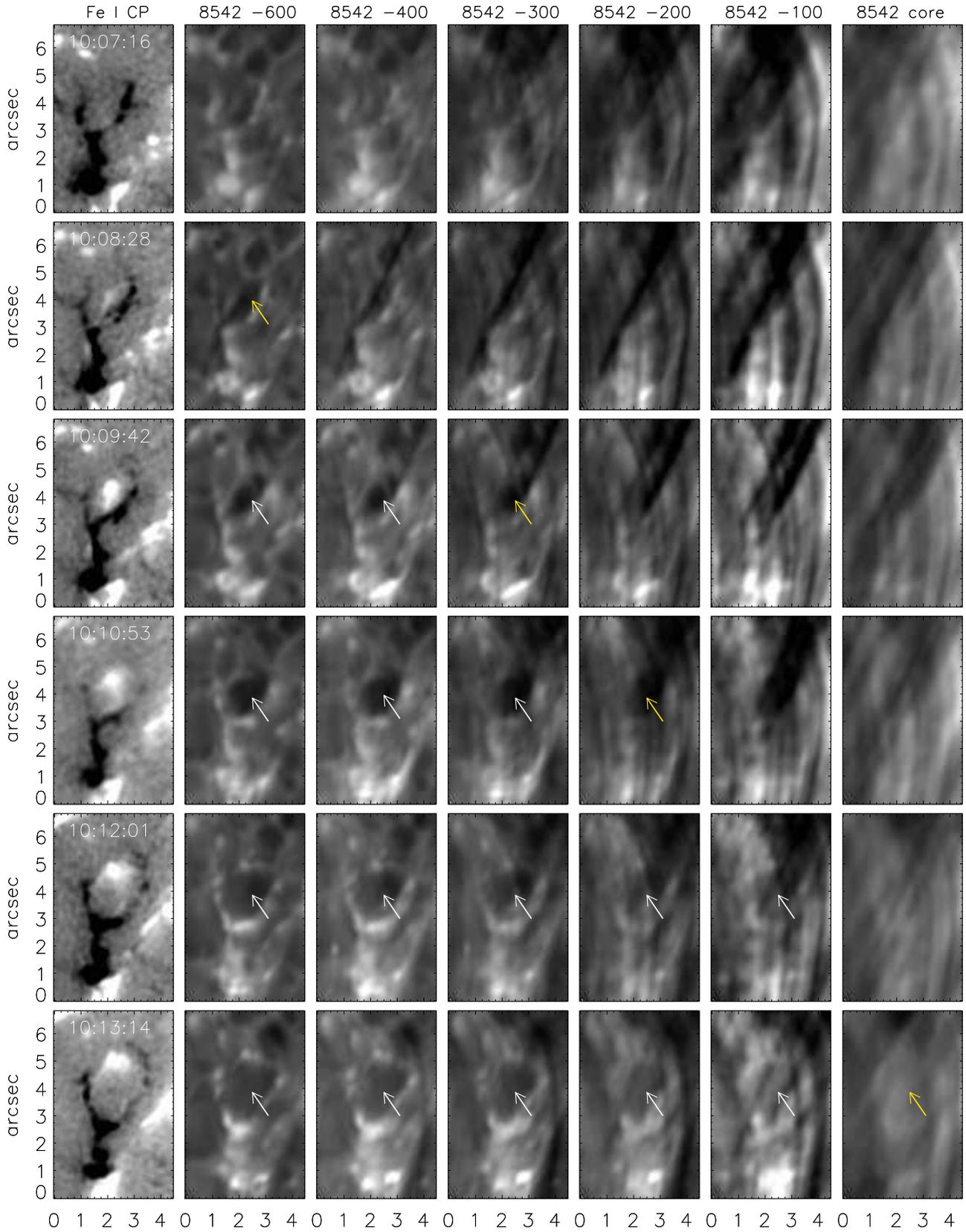}}
\caption{Close-up of the blue wing of the \ion{Ca}{2} 854.2~nm line,
  from 10:07:16~UT to 10:13:14~UT. The first column shows photospheric
  circular polarization maps. The next columns display \ion{Ca}{2}
  854.2~nm filtergrams at $-0.06$, $-0.04$, $-0.03$, $-0.02$, $-0.01$
  and $0.0$~nm from the line core.  White arrows indicate a clear
  detection of dark bubble \#1. {\bf Yellow arrows mean a more
    ambiguous detection, either because the bubble is overlaid by
    chromospheric fibrils or because it is not a conspicuous dark
    feature. In the former case, the presence of the bubble can be
    inferred from the brightenings surrounding it.}  
    \label{fig6}}
\end{figure*}

\begin{figure*}  
\centering  
\resizebox{0.85\hsize}{!}{\includegraphics{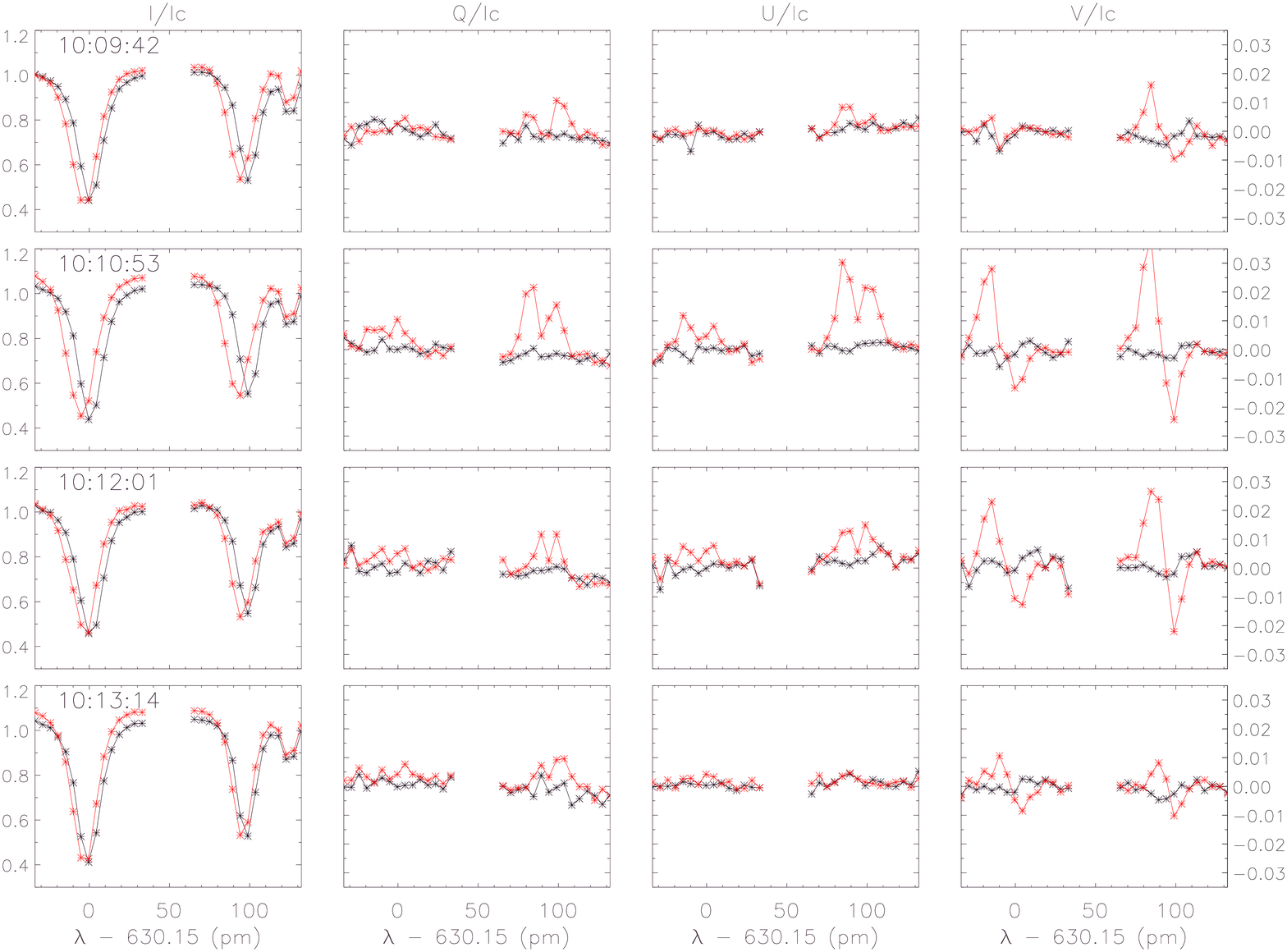}}  
\caption{Stokes profiles of the \ion{Fe}{1}~630.15 and 630.25~nm lines
  observed in the two pixels marked with yellow crosses in
  Figure~\ref{fig4}. They represent the dark bubble interior 
  (red lines) and pixels outside it (black lines). The 
  time of the scans is indicated in the upper left corner of each
  Stokes I panel. The selected line scans are the ones in which the
  dark bubble is more clearly visible (lower panels of
  Figure~\ref{fig2}) and in which significant LP signal is detectable
  (Figure~\ref{fig4}).
  \label{fig10}}
\end{figure*}

The evolution of the magnetic bubble is dominated by the separation of
the opposite magnetic polarities (column 3 of Figure~\ref{fig4}). The
magnetic bubble seems to leave footprints in the chromospheric
intensity images, and brightenings are seen at all wavelengths of the
\ion{Ca}{2} 854.2~nm scan, resembling the shape and position of the
magnetic features observed in the photospheric magnetograms.

To show how the dark bubble makes its way through the solar atmosphere
in more detail, Figure~\ref{fig6} presents a close-up of the blue wing
of the \ion{Ca}{2} 854.2~nm line. We only consider the time interval
during which the dark bubble exists, i.e., from 10:09:42~UT to
10:13:14~UT, plus two time steps before its appearance for
completeness. The first column shows maps of the photospheric circular
polarization to illustrate the presence and evolution of the magnetic
bubble. The next columns display, respectively, \ion{Ca}{2} 854.2~nm
filtergrams at $-0.06$, $-0.04$, $-0.03$, $-0.02$, $-0.01$, and $0$~nm
from line center.  Footprints of the magnetic bubble can be clearly
seen at all wavelengths in \ion{Ca}{2} 854.2 nm in the form of
brightenings.  White arrows point to the dark bubble when it is
unambiguously detected, while yellow arrows indicate a more vague
presence of the bubble.  In the latter cases, we do not directly
observe the dark bubble, but we infer its presence from the
circle-shaped brightenings that coincide with the magnetic legs. An
example can be found in the line core filtergram at 10:13:14~UT. As
the line core is approached, opaque chromospheric features (fibrils),
presumably located at greater heights, appear in the FOV and do not
allow one to see deep in the atmosphere.  In spite of this, we detect
the dark bubble at 10:09:42~UT in the \ion{Ca}{2} 854.2 $-0.06$~nm
filtergram and, one minute later, also in \ion{Ca}{2} 854.2
$-0.04$~nm. At 10:10:53~UT the bubble is still visible in those
images, but it has progressed towards the \ion{Ca}{2} 854.2 $-0.03$~nm
filtergram.  At that time, by chance, the chromospheric fibrils seem
to be on top of the dark bubble. By 10:12:01~UT the bubble has reached
the \ion{Ca}{2} 854.2 $-0.01$~nm filtergram, and at 10:13:14~UT it is
detected at all wavelengths, including the line core.  The behavior of
the bubble in the red wing (not shown here but in Figure~\ref{fig4})
is similar. The fact that the dark bubble approaches the core of the
\ion{Ca}{2} 854.2~nm line from the filtergram at $-0.06$~nm means that
it is traveling from the mid photosphere to the mid chromosphere over
a distance estimated to be about 1100 km (Figure~\ref{figcont}). Since
it takes some 215~s for the dark bubble to get there, its average
vertical speed is 5.2~km\,s$^{-1}$.

\subsection{Polarization profiles}
In this Section we discuss the properties of the observed Stokes
profiles before extracting more quantitative information from them.
 
\subsubsection{\ion{Fe}{1}~630~nm lines}

Figure~\ref{fig10} shows the \ion{Fe}{1}~630.15 and 630.25~nm Stokes
profiles emerging from two different pixels, one inside and the other
outside of {\bf dark bubble \#1} (red and black lines, respectively).
Those pixels have been marked with yellow crosses in
Figure~\ref{fig4}. Four time steps are displayed, corresponding to the
initial phases of the dark bubble development (10:09:42, 10:10:53,
10:12:01 and 10:13:14~UT). As can be seen, all intensity profiles are
blueshifted with respect to the pixel located outside of the magnetic
bubble region. The blueshift grows from 10:09:42 UT, attains a maximum
at 10:10:53 UT, and decreases in the following scans. This pattern
confirms that the magnetic bubble is indeed rising in the atmosphere.

\begin{figure*}
\centering
\resizebox{0.93\hsize}{!}{\includegraphics{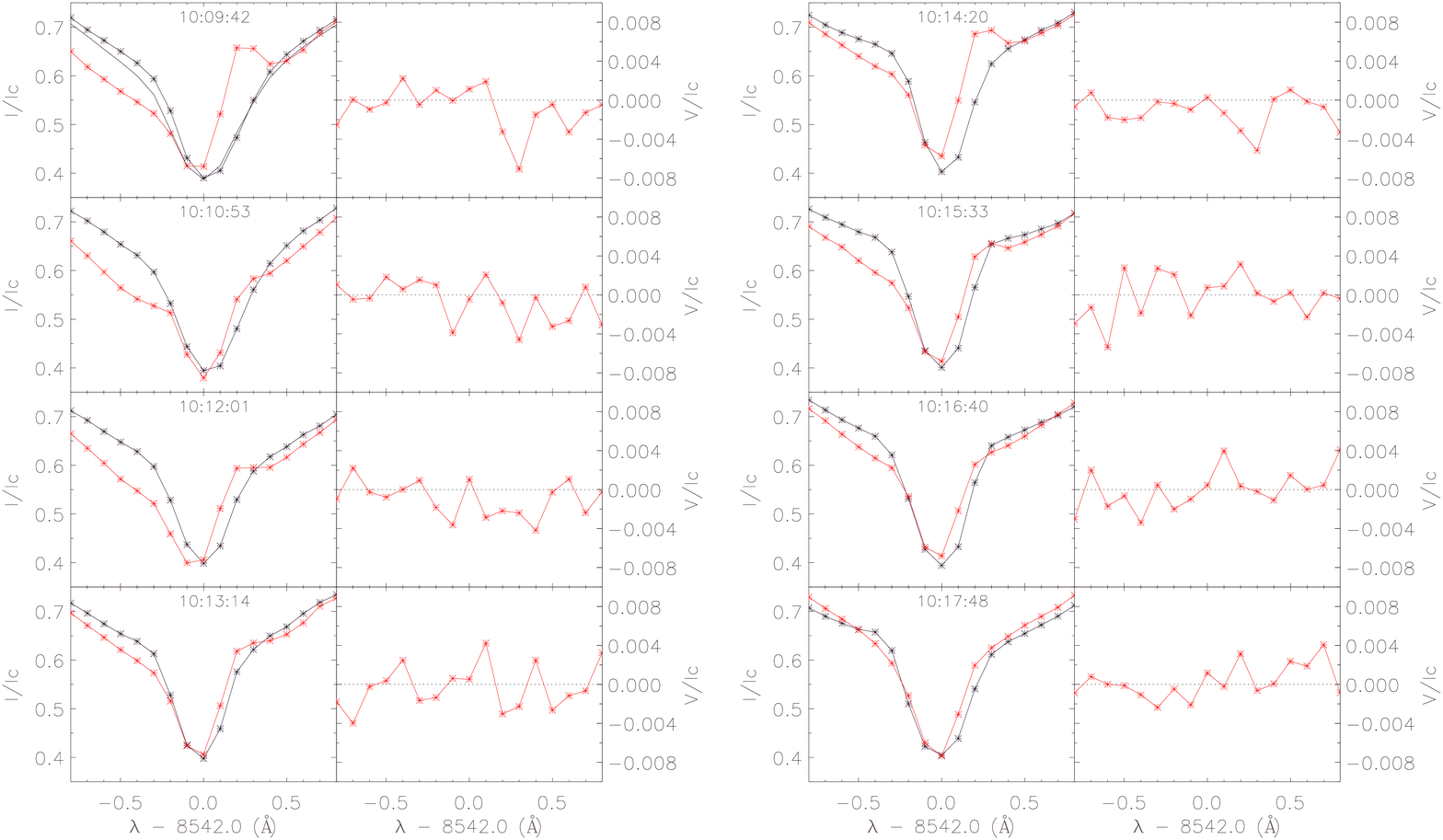}}
\caption{Temporal evolution of the \ion{Ca}{2} 854.2 Stokes I and V
  profiles observed in two pixels, one inside {\bf dark bubble \#1} (red
  lines) and another outside it (black lines).  Those
  pixels have been highlighted with yellow crosses in
  Figure~\ref{fig4}. A quiet Sun profile has been plotted for
  comparison in the first panel (simple black line).  Note the clear
  blueshifts of $-2.9$~km\,s$^{-1}$ at 10:09:42~UT and
  $-2.7$~km\,s$^{-1}$ at 10:12:01~UT.\label{fig9}}
\end{figure*}

The Stokes V profiles inside the emerging flux region show significant
signals of up to 3.5\%\,$I_{\rm c}$, meaning that during the four line
scans there was a magnetic field present in the bubble. They also
exhibit very strong blueshifts.  Stokes $Q$ and $U$, by contrast, have
large amplitudes only at 10:10:53 and 10:12:01~UT. This agrees with
Figure~\ref{fig4}, where those two line scans are the ones showing the
stronger and more extended LP patches.  A clear Stokes $Q$ signal is
detectable in the \ion{Fe}{1}~630.25~nm line at 10:09:42 and
10:13:14~UT, but with reduced amplitude, consistent with the idea that
at those particular times a horizontal magnetic field was entering and
then leaving the line formation region in the photosphere.

The inversion of the Stokes profiles of the pixel inside the dark
bubble yields field strengths of 480, 610, 360 and 230~G for each of
the time steps depicted in Figure~\ref{fig10}. The corresponding
inclinations reveal a horizontal field becoming slightly more
vertical: 87, 75, 72 and 70$^\circ$, respectively. Some single-lobed
Stokes V profiles are observed inside the dark bubble.  Jumps or
strong gradients of the physical parameters along the line of sight
can explain such kind of profiles \citep{alberto:2012}, suggesting
that a magnetic discontinuity is present in the atmosphere above the
bubble.

\subsubsection{\ion{Ca}{2}~854.2~nm}

Figure~\ref{fig9} shows the temporal evolution of the
\ion{Ca}{2}~854.2~nm Stokes I and V spectra observed in the two
pixels used above. An average quiet Sun profile is overplotted for
comparison (simple black line in the first panel).  Note that,
regardless of time, the intensity profiles coming from the pixel
outside the dark bubble and the quiet Sun are very similar.

Inside the dark bubble, the \ion{Ca}{2} 854.2~nm line shows enhanced
absorption in the blue wing. Some of the profiles also show an
emission peak (very prominent in particular at 10:09:42 and
10:14:20~UT).  The peak lies at the ``knee'' of the line---roughly
indicating an upper photospheric/low chromospheric origin---and occurs
only in the red wing, not in both wings.  This assymetry implies the
existence of very strong, localized upflows in layers below the
formation height of the line core, which move the opacity bluewards
and expose deeper atmospheric layers but do not significantly shift
the core itself.  The mechanism is similar to that described by
\citet{leenarts:2010} in the first quartet of their Figure 9 or by
\citet{leenarts:2009} in their Figure 5, although the flows are of
opposite direction here and probably less intense.
To explain the emission feature one may also need a strong
temperature enhancement, but this will be investigated in Paper II of
this series using non-LTE inversions.

The intensity profiles observed in the bubble are strongly blueshifted
at some particular times (10:09:42, 10:12:01 and 10:14:20~UT). Part of
the blueshift might be induced by the presence of the emission peak,
but part of it must be real because of the enhanced absorption
occurring in the blue wing of the line.

\begin{figure*} 
\centering 
\resizebox{\hsize}{!}{\includegraphics{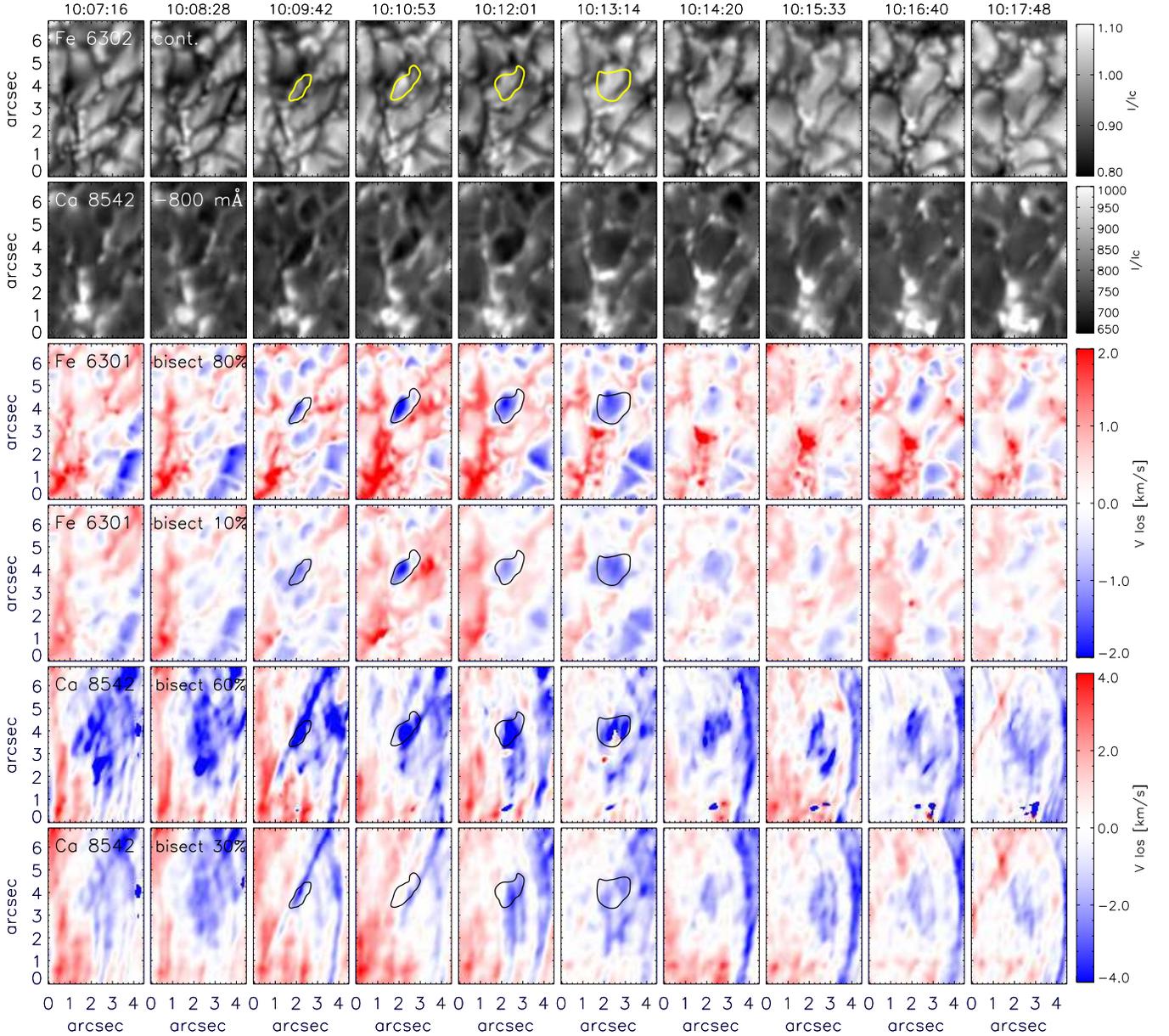}} 
\caption{Temporal sequence of intensities and LOS velocities for flux
  emergence case \#1, covering the interval from 10:07:16 to
  10:17:48~UT. First and second rows: continuum intensity at 630.32~nm
  and \ion{Ca}{2} 854.2 $-0.08$~nm intensity filtergrams. Third and
  fourth rows: \ion{Fe}{1}~630.15~nm bisector velocities computed at
  the 80\% and 10\% intensity levels. Fifth and sixth rows:
  chromospheric velocities determined from \ion{Ca}{2} 854.2~nm line
  bisectors at the 60\% and 30\% intensity levels.  Velocities are
  clipped at $\pm 2$ km\,s$^{-1}$ in the photosphere and $\pm
  4$~km\,s$^{-1}$ in the chromosphere. Note the strong redshifts of
  2~km\,s$^{-1}$ that appear in coincidence with the negative polarity
  foot, particularly at the 80\% intensity level around
  $(x,y)=(1\arcsec-2\arcsec, 2\arcsec-3\arcsec)$.  Brigthenings also
  occur at the same position.  The contours indicate the
  presence of the dark bubble during its life. \label{fig7}}
\end{figure*}

Outside of the magnetic feet, the dark bubble exhibits no
\ion{Ca}{2}~854.2~nm Stokes V signal above the noise level except at
the wavelength of the intensity bump produced by the emission peak.
Only there a weak chromospheric circular polarization signal can be
measured.  Although the profiles are very noisy, the existence of this
signal is undisputable because it produces extended patches in the
corresponding monochromatic Stokes~V images.  We will see in
Figure~\ref{fig11} that pixels with emission features are found mainly
at the edges of the magnetic bubble, with only a few lying inside.

\subsection{Velocity of the rising gas}

In this Section we discuss the dynamics of the emerging bubble based
on the bisector velocities derived from the observed Stokes~I profiles.

Figure~\ref{fig7} shows LOS velocity maps for flux emergence case \#1
from 10:07:16 until 10:17:48~UT.  The first and second rows display
continuum intensity maps at 630.32~nm and \ion{Ca}{2} 854.2 $-0.08$~nm
filtergrams, the third and fourth rows photospheric velocities at the
80\% and 10\% intensity levels, and the fifth and sixth rows
chromospheric velocities at the 60\% and 30\% levels.  The velocities
are clipped at $\pm 2$~km\,s$^{-1}$ for the photosphere and at
$\pm 4$~km\,s$^{-1}$ for the chromosphere. Negative velocities
indicate blueshifts. Black contours in the velocity panels 
indicate the position of the dark bubble.

In the photospheric {\bf velocity} maps (rows three and four), the
granulation pattern is clearly visible throughout the temporal
sequence. At 10:09:42~UT, a granule near the center of the
FOV with an elongated shape shows more prominent blueshifts of $-1.8$
and $-1.3$~km\,s$^{-1}$ at the 80\% and 10\% intensity levels,
respectively.  One minute later, at 10:10:53~UT, the blueshifts have
increased to $-2.4$ and $-2.0$~km\,s$^{-1}$, coinciding with the
stronger linear polarization patch. At 10:12:01~UT, the LOS velocities
start to decrease down to the $-1.8$ and $-1.1$~km\,s$^{-1}$ attained
at 10:13:14~UT.  Thus, the blueshifts increase their value at the
beginning, peak at 10:10:53~UT (regardless of the bisector level), and
decrease in the next line scans.  The bisector closer to the line core
reveals smaller Doppler shifts than the bisector closer to the
continuum.

\begin{figure*}
\centering
\resizebox{\hsize}{!}{\includegraphics{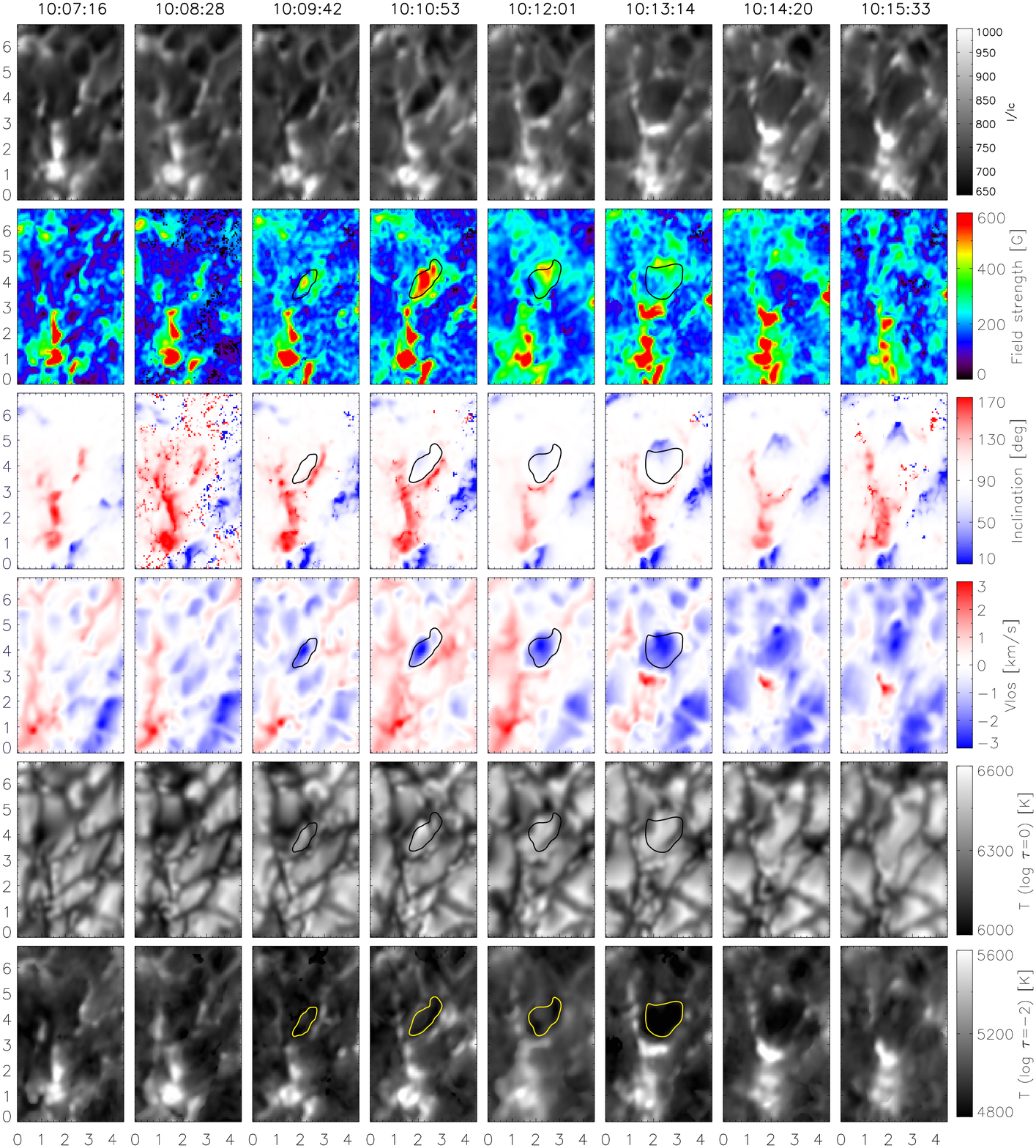}}
\caption{Temporal evolution of the atmospheric parameters inferred
  from the inversion of the \ion{Fe}{1}~630~nm lines {\bf for flux 
  emergence event \#1}. From top to
  bottom: \ion{Ca}{2}~854.2~nm $-0.08$ filtergram, field
  strength, inclination of the vector magnetic field, LOS velocity,
  temperature at $\log \tau=0$, and temperature at $\log \tau =-2$.
  The sequence covers the interval from 10:07:16 to 10:15:33~UT. 
  The contours outline the position of the dark bubble.
\label{fig13}}
\end{figure*}

Redshifts are also observed at the feet of the bubble in the
photospheric velocity panels. Their average value is 2~km\,s$^{-1}$ as
measured at the 80\% intensity level.  Peaks of 3~km\,s$^{-1}$ are
reached at 10:10:53 and 10:14:20~UT, while velocities of
3.7~km\,s$^{-1}$ are detected at 10:15:33~UT. These strong downward
flows are probably the result of drainage of material along the legs
of the rising magnetic bubble. They coincide with strong brightenings
in the intensity filtergrams around $(x,y) = (1\arcsec, 0\arcsec-3\arcsec)$ 
and $(2\arcsec,2\arcsec-3\arcsec)$. Interestingly, those locations are
associated with the negative polarity leg and present a strong
magnetic field. The 10\% bisector level also shows redshifts at the
same positions, but of lesser magnitude.

The last two rows of Figure~\ref{fig7} display chromospheric
velocities as measured by the \ion{Ca}{2} 854.2~nm bisectors. At the
beginning of the temporal sequence, most of the FOV is dominated by
strong blueshifts associated with chromospheric fibrils that are
opaque and do not permit the detection of features below them. At
10:09:42~UT a narrow fibril seems to cross over the dark bubble, with
an average velocity of $-5.0$~km\,s$^{-1}$.  In the next time step the
fibril has mostly disappeared and the 60\% bisector velocity shows
blueshifts of $-5.5$ km\,s$^{-1}$ at the position of the dark bubble.
The bubble is well detected until the end of the sequence when the
velocities have slowed down to $-1$~km\,s$^{-1}$.  The 30\% bisector
velocity indicate weaker upward motions in the dark bubble with
signals that are delayed with respect to the 60\% bisector velocity.
At those heights, the fibrils are detected until 10:10:53~UT, giving
way to the bubble at 10:12:01~UT with blueshifts of $-3$~km\,s$^{-1}$.
The velocity slows down to $-1$~km\,s$^{-1}$ at the end of the
sequence.

As mentioned before, there is the possibility that some of the
blueshifts inferred from the \ion{Ca}{2} 854.2~nm line are not real
Doppler velocities but an artifact of the emission peak that occurs in
the red wing of the line at some locations. However, most of the
bubble interior does not show profiles in emission (see the blue
contours in Figure~\ref{fig11}), so a large fraction of the blueshifts
detected in the chromosphere must be legitimate. This issue will be
examined in more detail in Paper II of this series.

\subsection{Magnetic properties}

\subsubsection{Photospheric fields}

From top to bottom, Figure~\ref{fig13} presents the temporal evolution
of the \ion{Ca}{2}~854.2 $-0.08$~nm filtergrams, together with the
field strength, field inclination, LOS velocity, and temperature at
$\log \tau =0$ and $\log \tau=-2$ inferred from the inversion of the
Stokes profiles of the two \ion{Fe}{1} lines, {\bf for flux emergence
  event \#1}. The appearance of the positive leg is well visible at
10:09:42~UT in the field strength map and one minute later in the
inclination map. It then builds up. The newly formed positive leg
contains field strengths that vary between 300 and 500~G from its
appearance at 10:09:42~UT until the end of the sequence. Inclinations
change from 86$^\circ$ when the leg appears to 50$^\circ$ at the end
of the sequence. The negative leg yields stronger fields, ranging from
400 to 800~G, and inclinations varying between 145$^\circ$ and
110$^\circ$. The pattern of the field strength at the center of the
magnetic bubble follows that shown in Figure~\ref{fig4} by the linear
polarization patch. At 10:09:42 UT the center of the bubble presents a
magnetic strength of 300~G which grows up to 480 G in the next time
step and then decreases to 370, 200 and finally 120~G at 10:14:20 UT.
The inclinations at the center of the bubble lie between 75 and
95$^\circ$. Thus, a horizontal field traverses the photosphere being
particularly strong at 10:10:53 and 10:12:01 UT, in agreement with
Figure~\ref{fig4}.

The velocities show a normal granulation pattern in the first frames,
but later on, the granule at the center of the FOV stands out among
the others with increased upward velocities which reach a peak of
$-3$~km\,s$^{-1}$ at 10:10:53~UT before slowing down to
$-1$~km\,s$^{-1}$ at 10:15:33~UT.  The photosperic temperature at
$\log \tau=0$ does not show any trace of the dark bubble, in agreement
with the observed continuum intensity pattern. However, the dark
bubble is easily identified in the temperature map at $\log \tau=-2$,
where it shows a time-averaged deficit of 150~K with respect to the
mean temperature of 5020~K. At 10:13:14 UT the deficit reaches its
maximum value of 250~K. Note that the dark bubble seen in the
temperature maps cannot be mistaken with reverse granulation, since
its size is considerably larger than that of the underlying granule.
This means that the magnetic bubble is cooler than the surroundings in
the mid photosphere, but not below (we recall that $\log \tau = 0$
refers to the continuum forming region and $\log \tau = -2$ to a layer
some 300~km above it.)

\subsubsection{Chromospheric fields}
\label{ret}

\begin{figure*}
  \centering
  \resizebox{\hsize}{!}{\includegraphics{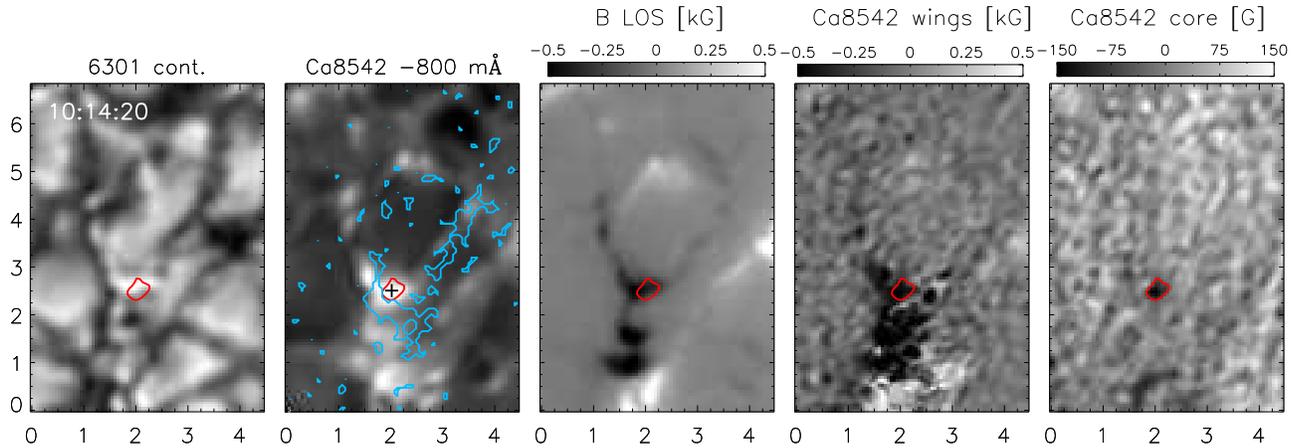}}
  \caption{From left to right: Stokes I map in the continuum at
    630.32~nm, \ion{Ca}{2} 854.2 $-0.08$~nm filtergram, longitudinal
    photospheric magnetic field from the inversion of the two
    \ion{Fe}{1} lines, and longitudinal magnetic field from the wings
    and the core of the \ion{Ca}{2} 854.2~nm line {\bf for flux
      emergence event \#1}. All the maps correspond to 10:14:20~UT.
    The red contours highlight regions with chromospheric longitudinal
    fields stronger than $-90$~G, and have been superimposed onto the
    other four panels. The cross in the second panel marks a pixel in
    the magnetic legs which is discussed in the text.  The blue
    contours in the second panel outline pixels with intensity
    profiles in emission. \label{fig11}}
\end{figure*}

The chromospheric magnetic signals are very small and thus difficult
to measure. We do not observe Stokes Q or U signals in the \ion{Ca}{2}
854.2~nm line above the noise level.  However, we do detect Stokes V
signals that allow us to estimate the longitudinal field in
chromospheric layers. We have utilized the weak field approximation
for this purpose, as described in Sect.~\ref{mag}.

Figure~\ref{fig11} shows intensity filtergrams in the low and mid
photosphere (first two panels), together with longitudinal magnetic
field maps in the low photosphere (panel 3), the mid photosphere
(panel 4), and the mid chromosphere (panel 5). The longitudinal field
map displayed in panel 3 has been constructed using the inversion
results as $B_{\rm LOS} = B \cos \gamma$. The longitudinal fields
derived from the \ion{Ca}{2} line have been obtained by applying the
weak field approximation to the line wings (excluding the central $\pm
300$~m\AA; panel 4) and to the region within the profile 'knees' ($\pm
300$~m\AA\ around the line core; panel 5), in order to separate the
photospheric contribution from the purely chromospheric contribution.

In Figure~\ref{fig11}, the red contours outline pixels with
chromospheric longitudinal fields stronger than $\phi = -90$ G (three
times the noise level). As can be seen, the most intense chromospheric
fields occur at the position of the lower half-moon-shaped magnetic
leg. The pixel marked with a cross in panel 2 at
$(x,y)=(2\arcsec,2.5\arcsec)$, for example, has a field strength of
750~G and an inclination of 123$^\circ$ in the photosphere, i.e., a
longitudinal field of $-410$~G, but only $-175$~G in the mid
chromoshere as inferred from the \ion{Ca}{2} 854.2~nm line core. This
change by a factor of 2.3 between the two layers sampled by the
corresponding maps implies a rapid decrease of the field strength with
height or a progressively larger inclination of the field towards
higher layers, or both.

\begin{figure*}
\centering
\resizebox{\hsize}{!}{\includegraphics{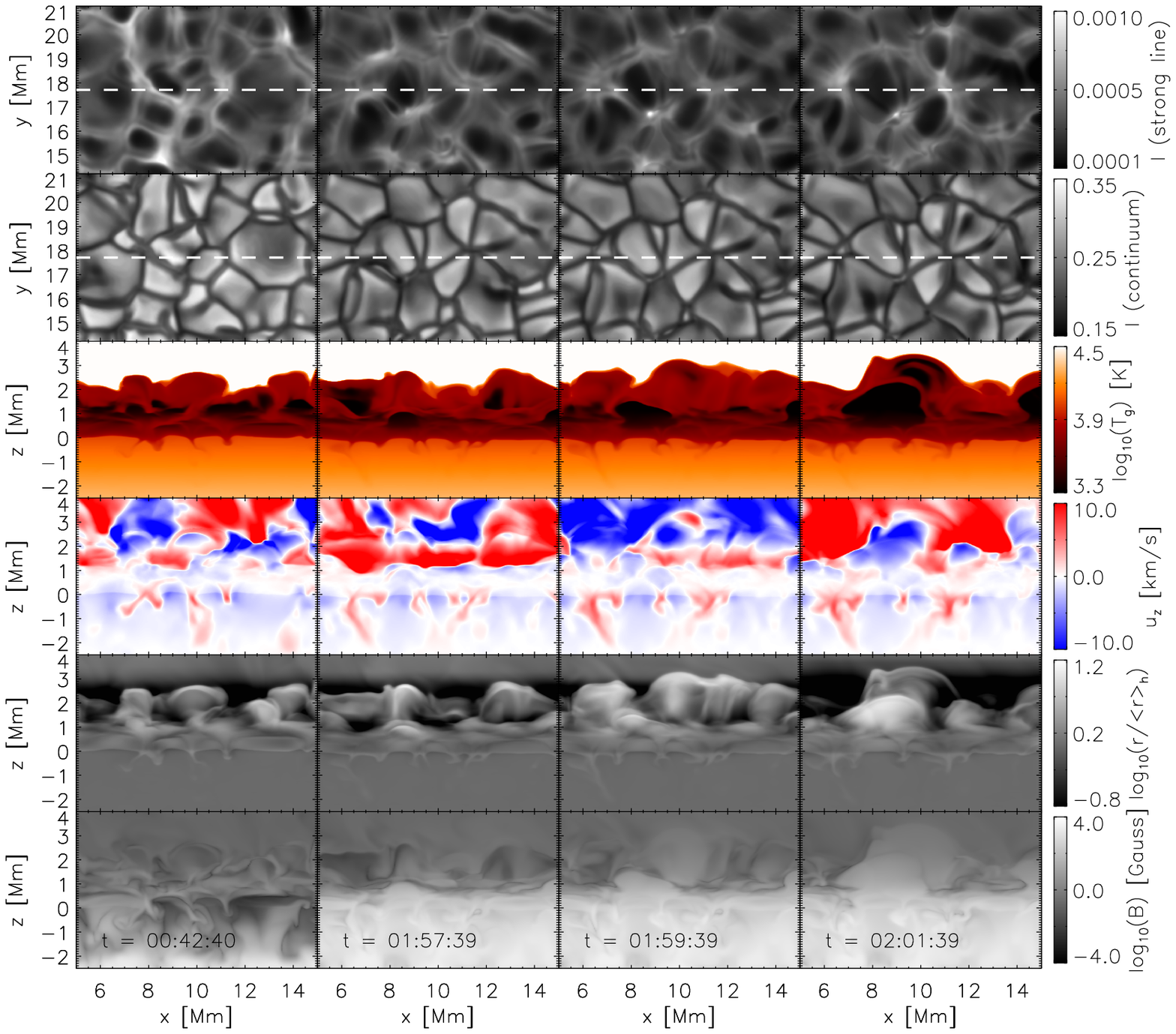}}
\caption{Views of the simulation box at four different times.  {\bf
    Top two rows: intensities in the horizontal plane that would be
    observed in the core of a strong chromospheric line (such as
    Ca~{\sc ii} 854.2~nm) and in the continuum at 630~nm. Lower four
    rows: vertical cuts showing temperature, vertical velocity, gas
    density, and magnetic field strength. The dashed lines in the
    first two rows indicate the position of the selected vertical
    slices}.
\label{fig:emergence_total_iy=372}}
\end{figure*}

The polarity of the magnetic feet is the same---negative---in the
photosphere and the chromosphere. But that does not always seem to
hold for pixels in the {\em vicinity} of the magnetic feet, as some of
them show {\em opposite-polarity} (positive) V profiles in the
chromosphere. We have pinpointed those pixels with blue contours in
the second panel of Figure~\ref{fig11}. Whenever there is a change in
the sign of Stokes~V between the photosphere and the chromosphere, the
\ion{Ca}{2} 854.2~nm Stokes I profiles show emission instead of
absorption. This reversal of the polarity is only apparent: in the
weak field regime, the Stokes V sign is determined by the sign of
$\delta I / \delta \lambda$.  When the line goes in emission, $\delta
I / \delta \lambda$ changes sign and so does Stokes V. Thus, also for
those pixels the magnetic polarity is the same in the two atmospheric
layers.  We have previously observed this behavior in the Stokes~I and
V profiles of Figure~\ref{fig9}, which switch from absorption to an
assymetric emission peak in the red wing. Other examples of apparent
polarity reversals in chromospheric lines have been presented (and
explained) by \citet{val}, \citet{almeida}, and \citet{jaime:2013}.

The blue contours show that the magnetic bubble is surrounded by
pixels with chromospheric Stokes I profiles partially in emission,
which is indicative of plasma being displaced at a different speed and
possibly with much higher temperature in some layers along the LOS.
Whether such higher temperatures would be achieved by reconnection or
by some other heating mechanisms remains to be determined.

\section{Numerical simulations}
\label{mhd}

\begin{figure*}
\centering
\resizebox{.8\hsize}{!}{\includegraphics{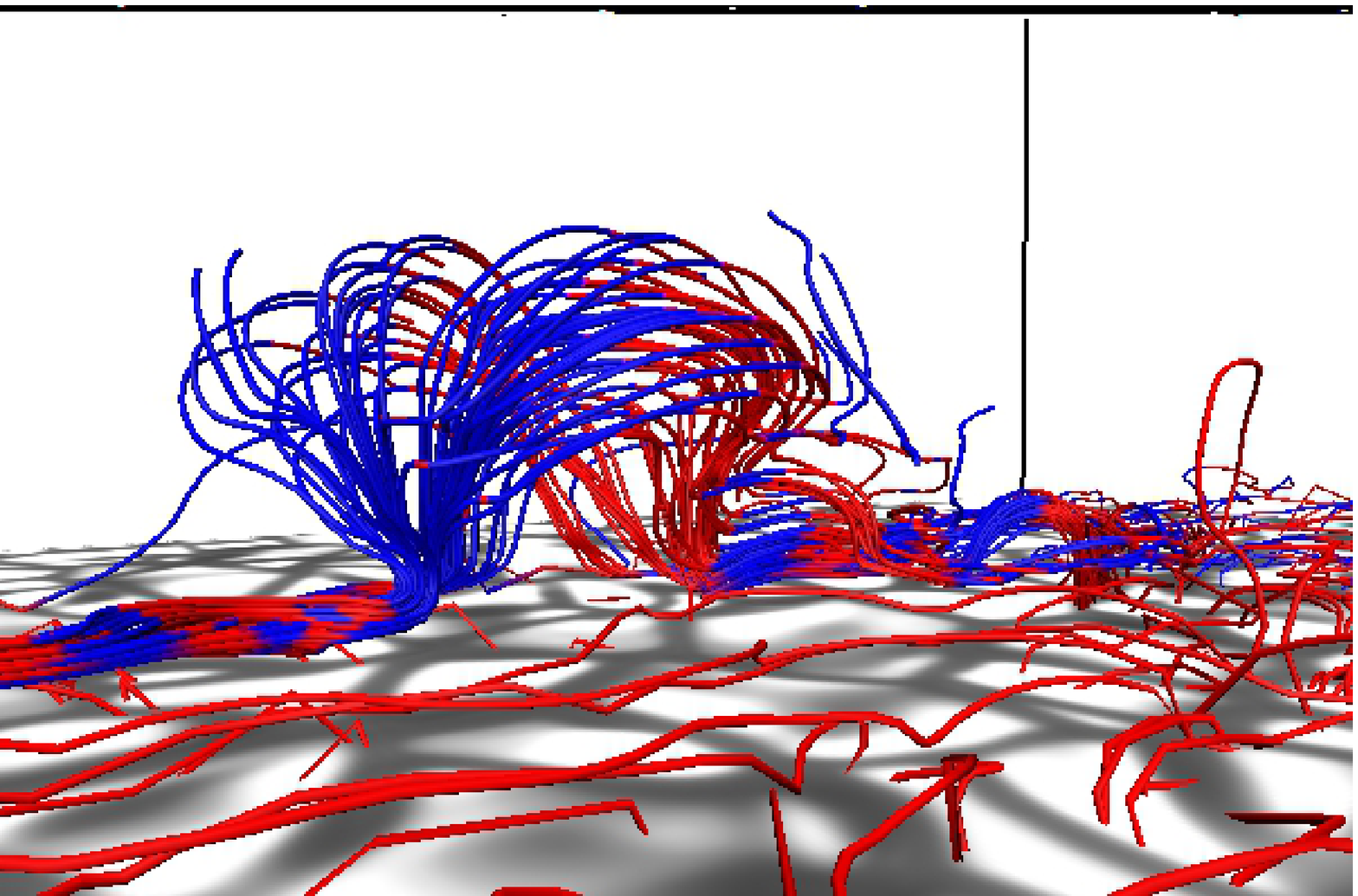}}
\vspace{0.5cm}
\caption{3D rendering {\bf extracted from the simulation} of the 
magnetic field lines producing the magnetic bubble. {\bf The magnetic bubble reaches a diameter of 3Mm}. Red and blue 
indicate fields pointing upward and downward, respectively.  
\label{vapor}}
\end{figure*}

Further insight into the observed phenomena may be gained with the
help of `realistic' numerical simulations.  In this case the term
`realistic' means models in which enough physical mechanisms are
included so that observables synthesized from the simulations can be
directly compared with our observations. We have run an experiment in
which a magnetic flux sheet that is injected into the convection zone
rises to the photosphere and later breaks through into the
chromosphere in much the same way as we believe is happening in the
observations described above.

The models presented here are performed on a $24\times 24\times
16.5$~Mm$^3$ cube discretized on a grid containing $504\times
504\times 496$ cells, with a horizontal grid size of $41$~km and a
variable spacing in the vertical direction ranging from $20$~km in the
photosphere, gradually increasing to $95$~km in the corona 14~Mm above
the photosphere and to $85$~km in the convection zone 2.5~Mm below the
photosphere. The model is calculated using the Bifrost code
\citep{2011A&A...531A.154G} and includes an equation of state based on
solar abundances in LTE, optically thick radiative transfer and losses
from the photosphere and chromosphere including scattering
\citep{2000ApJ...536..465S,hayek2010}. In the middle and upper
chromosphere as well as in the transition region, a recipe for
effectively thin radiation is used, as described by
\citet{2012A&A...539A..39C}. In the corona, thermal conduction along
the magnetic field plays an important role; this is treated by
operator splitting where the resulting implicit operator for
conduction is solved using a multi-grid method as described in
\citet{2011A&A...531A.154G}.

In the simulation a magnetic sheet aligned with the $y$-axis,
stretching from $x=4$~Mm to $x=16$~Mm, and with strength $B_y=3363$~G
is inserted at the bottom boundary. {\bf This particular field
strength has been chosen after experimenting with weaker flux sheets
that failed to produce the desired emergence}. The insertion continues
for 1~hour 45~minutes (6310~s) before it is turned off. The magnetic
field of the sheet reaches the photosphere fairly rapidly (within an
hour or so), but at the photosphere is no longer carried by convection
nor sufficiently buoyant and therefore becomes `stuck' slowly gaining
strength as field from below piles up.

\begin{figure*}
\centering
\resizebox{0.85\hsize}{!}{\includegraphics{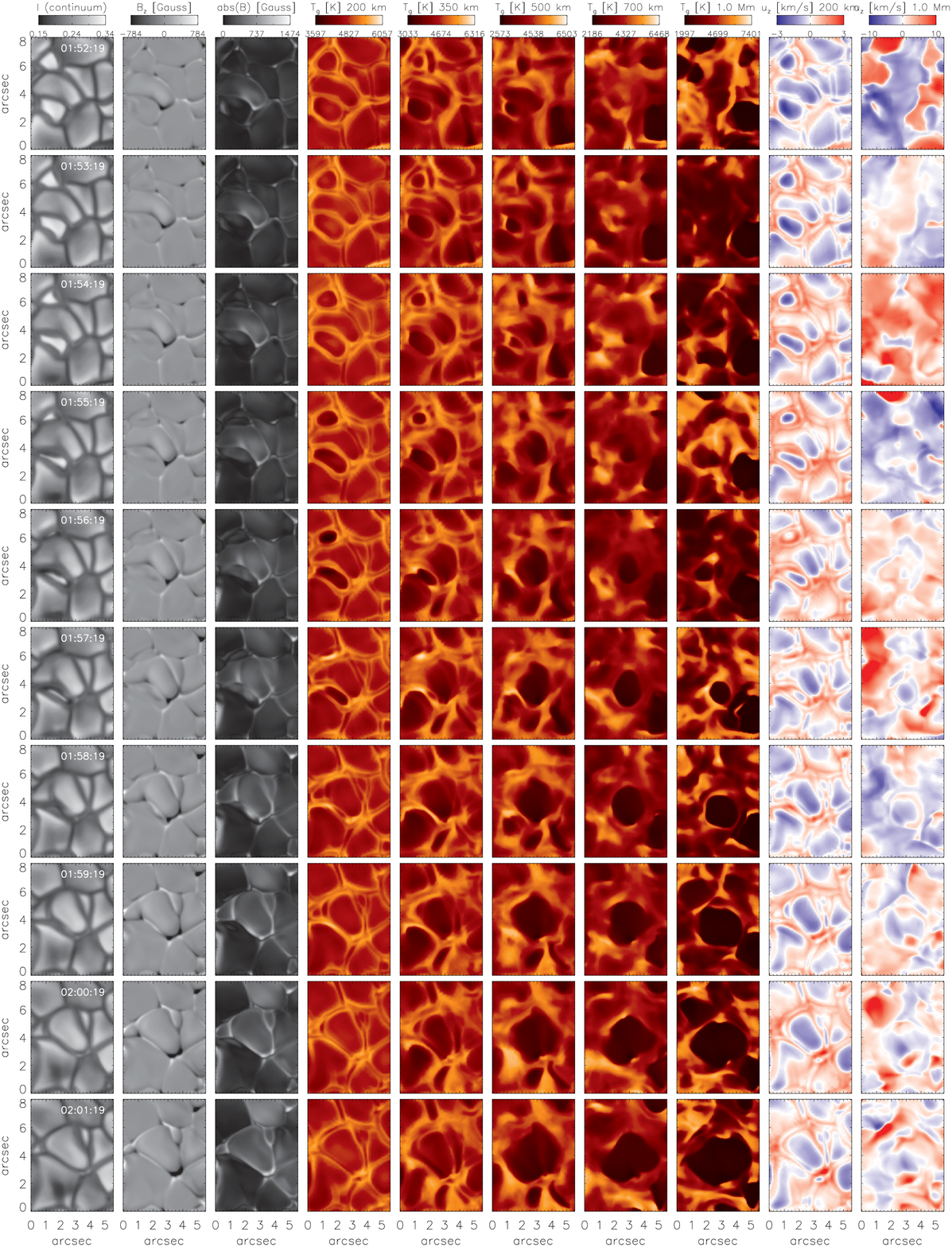}}
\caption{Temporal evolution between 1:52:19 and 2:01:19 simulation
  time of the white light intensity, vertical magnetic field, total
  magnetic field, temperature at 200, 350, 500, 700, 1000~km above the
  photosphere, as well as the vertical velocity at 200 and 1000~km
  above the photosphere. The color scale for $B_z$ is set to
  $[-784,784]$~G, while for $\mid\! B\!\mid$ is set to $[0,1474]$~G.
  The temperature scale ranges between 3597 and 6057~K at 200~km, 
  between 3033 and 6316~K at 350~km,
  between 2573 and 6503~K at 500~km, betweeen 2186 and 6468~K at
  700~km, and between 1997 and 7401~K at 1000~km. The velocity scale at
  200~km is set to $[-3,3]$~km\,s$^{-1}$ and at 1000~km to
  $[-10,10]$~km\,s$^{-1}$ (red color represents downflows).\label{fig12}}
\end{figure*}

In Figure~\ref{fig:emergence_total_iy=372} we show a vertical cut of
the evolution of the magnetic field strength, density, vertical
velocity, and temperature along with the emergent intensity from the
photosphere and the lower chromosphere (i.e., similar to the emergent
intensity in the white light continuum and near the line core of a
strong chromospheric line such as Ca~{\sc ii} 854.2~nm) at 4 different
times during the simulation. Initially the emerging magnetic field has
only reached to some 1.5~Mm below the photosphere, and the photosphere
and chromosphere are dominated by granular flows and chromospheric
oscillations. At $t=$ 1h 59m 39s, the field fills the convection zone
up to the photosphere, but has not yet penetrated the upper
photosphere nor the chromosphere. This happens first two minutes
later, when the field gradient in the vertical direction becomes
strong enough to trigger the magnetic buoyancy instability (Parker
instability): when the magnetic field strength decreases with height
it reduces the gas pressure gradient and sets up a situation where
heavier fluid overlies a lighter fluid as described by
\citet{2004A&A...426.1047A} and references therein. At this
point the field breaks through the photosphere and a bubble of high
field strength emerges into the chromosphere and corona. This
bubble expands rapidly and adiabatic cooling ensues; the chromospheric
material becomes very cool as the bubble expands far beyond the
original dimensions of the flux emerging into the photosphere. Though
the event shown here is too weak to significantly perturb the
granulation pattern in the photospere we see that, as the bubble
rises, the chromospheric emission becomes very low where the
temperatures are lowest, while enhanced emission occurs near the edges.
{\bf Upon expanding into the chromosphere, the bubble achieves a
  nearly spherical shape as can be seen in the 3D rendering of
  Figure~\ref{vapor}.  At this stage it has a diameter of some 3~Mm}.

The response of the simulated photosphere and chromosphere to the
expansion of the emerging magnetic flux is further shown in
Figure~\ref{fig12}. The evolution of the photospheric continuum
intensity and magnetic field are plotted alongside the temperature at
five separate heights ($200, 350, 500, 700, 1000$~km) in the lower and
middle chromosphere and the vertical velocity at two heights ($200,
1000$~km). Initially, a small region of strong magnetic field (both
vertical and horizontal) pierces the photosphere. At this time there
is no response in the temperature nor in the velocity pattern at
higher altitudes. One minute later, at 1:53:19, a circular region of
slightly cooler gas appears over the emerging flux region 350~km above
the continuum forming layer. (Note that we do not see any obvious
response in the temperature at a height of 200~km at that time, nor
indeed at any later time). This structure is similar to that appearing
in the inverted temperature maps at $\log \tau=-2$, which show the
dark bubble in the middle photosphere. The cool region at 350~km grows
larger as time progresses, and at 2:01:19 has a diameter of roughly
3\arcsec.  This temperature response is mirrored at greater heights,
but the cooling region appears later, at 1:54:19 at 700~km and at
1:55:19 at 1000~km, implying an ascent speed of 4.2~km\,s$^{-1}$.  At
these great heights the temperature becomes very low in the `bubble'
falling to less than 2000~K. Though the temperature response is very
weak or missing at 200~km the flux emergence is quite evident in the
velocity signal, with upflows of order 3~km\,s$^{-1}$ outlining the
cool bubble above.  At the edges of the granule, and especially
centered on the regions of greater vertical field strength, there are
downflows of 3~km\,s$^{-1}$.  At 1000~km, where the bubble is very
pronounced in temperature, we find upflows of some 10~km\,s$^{-1}$
surrounded by downflows with approximately the same amplitude.

In the photosphere the horizontal magnetic field strength of the
rising flux is of the order several hundred~G, increasing to 1~kG near
the edges of the granule containing the emerging field. The vertical
field is much more concentrated to the intergranular lanes, and
reaches a maximum of some $1150$~G. At a height of 700~km we find
that, as the magnetic bubble rises and expands, the horizontal
magnetic field grows to $B_{\rm h}=100$~G and is concentrated towards
the center of the bubble, and while the vertical field grows to a
similar amplitude, it is concentrated to the edges.

The evolution of the simulated dark bubble is, with the exception of
perturbed granulation, remarkably similar to what is observed.  This
includes the field strength and LOS velocities of the emerging flux in
the photosphere and the chromosphere, the ascent speed and expansion
velocity of the bubble, its temperature deficit in the middle
photosphere, and the duration of the entire event.  In the model, the
emerging field continues to expand and eventually, after a series of
reconnection events, merges and/or replaces the ambient coronal field.

\section{Discussion and conclusions}
\label{disc}

With the advent of very high resolution spectropolarimetric
measurements, our understanding of small-scale flux emergence in the
solar atmosphere has improved dramatically. So has the realism of
numerical simulations of this phenomenon, thanks to ever-increasing
computer capabilities.  Both within active regions and in the quiet
Sun, a common form of flux emergence is the rise of simple
$\Omega$-shaped magnetic loops that intersect the photosphere at two
magnetic footpoints.  The present paper deals with a more complex kind
of flux emergence, a 3D semi-sphere that we have called magnetic
bubble. Such a bubble pierces the photosphere producing half-moon
shaped magnetic feet.  To our knowledge, only \citet{salvo12} have
observed flux emergence events with similar crescent-shaped magnetic
legs. Here we present multiwavelength observations and simulations
that allow us to get a better understanding of the emergence of these
magnetic features.

We have observed some of the typical signatures of flux emergence:
abnormal granulation, separation of magnetic feet of opposite
polarity, and brightenings in chromospheric layers. We have quantified
how the magnetic bubble expands horizontally and rises vertically. The
expansion and rise seem to happen in an intermittent way, rather that
in a continuous and smooth fashion. But the most remarkable
observational feature---and the main result of this paper---is the
appearance of a dark bubble contained within the magnetic enclosure in
the mid-photosphere and above. This bubble is magnetized, and pumps
magnetic flux into the chromosphere.  Surrounding the bubble we have
detected \ion{Ca}{2} 854.2~nm intensity profiles with an emission peak
in their red wings.  In order to produce such a peak, a very high
velocity gradient is needed along the LOS, possibly coupled with
strong temperature enhancements at some height in the atmosphere.
These are perhaps the signs of reconnection happening at the edges of
the bubble, or of other heating mechanisms.

We have measured the distance between the magnetic feet and found
values of up to 6.6\arcsec, with horizontal separation velocities of
about 4~km\,s$^{-1}$ on average.
\citet{ot07} reported separation speeds of 4.2~km\,s$^{-1}$ during the
initial phases of flux emergence and some 1~km\,s$^{-1}$ later on.
The dark bubble shows maximum sizes of $3\arcsec \times 1\arcsec$ and
horizontal expansion velocities similar to those of the magnetic feet.

The rising gas has LOS velocities of around $-2.5$~km\,s$^{-1}$ in the
photosphere and $-3$~km\,s$^{-1}$ in the chromosphere.  Strong
downflows are detected at the position of the magnetic feet, with
peaks of 4~km\,s$^{-1}$. Observational studies like \citet{salvo08}
and \citet{salvo12} report slightly smaller values of $-1$ to
$-2$~km\,s$^{-1}$ for the upflows and 1 to 2 km\,s$^{-1}$ for the
downflows. Numerical simulations by \citet{mark08} yield upward
velocities as small as $-0.5$ km\,s$^{-1}$. However, the simulations
of \citet{juan08} give photospheric values of $-2.5$~km\,s$^{-1}$ for
the upflows and 3-5 km\,s$^{-1}$ for the downflows.  In the
chromosphere they find $-7~$km\,s$^{-1}$ and 10 km\,s$^{-1}$,
respectively. These values are similar to ours.

Brightenings, in particular where the new flux system interacts with
the preexisting ambient field, are common in emergence events as
reported by, e.g., \citet{salvo08} and \citet{vargas12}. We also
observed them.  Figure~\ref{fig4} shows that brightenings are seen all
across the \ion{Ca}{2} 854.2~nm line profile, resembling the shape and
position of the magnetic feet.

We have derived the magnetic properties of the emerging flux in the
photosphere and the chromosphere. The longitudinal magnetic field at
the position of the half-moon shaped magnetic legs is around 410~G in
the photosphere (corresponding to a field strength of 750~G) and 175~G
at the height of formation of the \ion{Ca}{2} 854.2~nm line core,
i.e., the middle chromosphere. The field strength in between the two
magnetic feet is around 300 to 400~G as determined from the inversion
of the observed \ion{Fe}{1} 630~nm lines.  \citet{salvo12} also
obtained maximum strengths of 400~G in the photosphere with the
IMAX instrument on board the SUNRISE balloon \citep{imax}. The
numerical simulations performed here yield hG strengths for the
horizontal field of the bubble interior and 1 kG for the vertical
field of the feet, 200~km above the continuum forming layer.  At a
height of 700~km the horizontal magnetic field has a strength of some
100~G and is concentrated towards the center of the bubble, whereas
the vertical field has similar strengths but is mainly seen near the
magnetic legs.  This is in excellent agreement with \citet{mark08} and
\citet{tor09}, who found $B_{\rm h}=100$~G at a height of 1000~km.
\citet{juan08} also obtained 80~G at 900~km.

Our realistic numerical simulations are able to reproduce, both
qualitatively and quantitatively, the main observational results.
This includes the field strength and the LOS velocities of the
emerging flux, both in the photosphere and in the chromosphere, the
expansion and rise of the bubble, the lower gas temperature of the
bubble in the mid and high photosphere, and the total lifetime.  The
organization of the field lines in the simulation is shown in
Figure~\ref{vapor}. As can be seen, this is a truly spherical bubble
which is expanding into the chromosphere after rising through the
photosphere. Except at the position of the bubble, the photospheric
field is essentially horizontal. The picture that can be drawn from
both observations and simulations is one of an extended sheet
of horizontal fields initially located in subphotospheric layers that
rise into the atmosphere with the help of the vertical upflows of a
granule. During the ascent, the fields adopt the form of a bubble and
create a temperature deficit which is observed as a dark feature
in the middle photosphere ($\log \tau = -2$) and above.

\citet{juan08} and \citet{tor09} detected similar dark bubbles in
their simulations. The bubbles are formed by adiabatic cooling of flux
emerging and expanding rapidly.  Both papers quote temperatures of
less that 3000~K for the dark bubbles at heights of 700 to 900~km.

Using Hinode measurements, \citet{vargas12} reported the existence of
small-scale (2-4 Mm), short-lived (12 min) dark areas in a flux
emergence event similar to the ones analyzed here. These features 
were observed in \ion{Ca}{2}~H filtergrams, but also in photospheric
\ion{Na}{1} 589.6~nm Stokes~I images.  According to \citet{vargas12},
they appear in regions with negligible longitudinal magnetic field.
Even though there are some differences (our bubbles present both
horizontal and vertical field and occur above $\log \tau = -2$), their
observations may have targeted the same features at less spatial
resolution and only with partial magnetic field information.

{\bf In Paper II of this series} we will invert the observed
\ion{Ca}{2}~854.2~nm profiles under non-LTE conditions to determine
the properties of the magnetic bubble in the chromosphere, including
the stratification of the gas temperature and the LOS velocity.  We
will investigate the origin of the emission features detected in the
\ion{Ca}{2}~854.2~nm red line wing and the possibility that they
signal the presence of strong temperature enhancements at some height
in the atmosphere.  Also, we will compute synthetic \ion{Fe}{1} 630~nm
and \ion{Ca}{2}~854.2~nm Stokes profiles from the simulations in order
to compare them with the observed ones.  Differences betweeen the
synthetic and observed profiles will provide additional constraints to
our numerical model of flux emergence.

\acknowledgments

Part of the work presented here was done while one of us (A.O.) was a
Visiting Scientist at the Instituto de Astrof\'{\i}sica de
Andaluc\'{\i}a (CSIC). Financial support by the Spanish MINECO through
project AYA2012-39636-C06-05, including a percentage from European
FEDER funds, by the Research Council of Norway through grants
208027/F50 and 'Solar Atmospheric Modelling', by the European Research
Council under the European Union's Seventh Framework Programme
\mbox{(FP7/2007-2013)} / ERC Grant agreement no.\ 291058, and by the
Programme for Supercomputing of the Research Council of Norway through 
grants of computing time are gratefully acknowledged. The Swedish
1~m Solar Telescope is operated by the Institute for Solar Physics of
Stockholm University in the Spanish Observatorio del Roque de los
Muchachos of the Instituto de Astrof\'{\i}sica de Canarias. This
research has made use of NASA's Astrophysical Data System.

\clearpage



\end{document}